\documentclass[preprintnumbers,amsmath,amssymb]{revtex4}

\usepackage{graphicx}
\usepackage{dcolumn}
\usepackage{bm}

\begin{document}

\title{Duality symmetries behind solutions of the classical simple pendulum}

\author{Rom\'{a}n Linares Romero} 
\email{lirr@xanum.uam.mx}

\affiliation{Departamento de F\'{\i}sica, Universidad Aut\'onoma Metropolitana Iztapalapa,\\
San Rafael Atlixco 186, C.P. 09340, M\'exico D.F., M\'exico,
}


\begin{abstract}
The solutions that describe the motion of the classical simple pendulum have been known for very long 
time and are given in terms of elliptic functions, which are doubly periodic functions in the complex plane.
The independent variable of the solutions  is time and it can be considered either as a real variable or as a purely 
imaginary one, which introduces a rich symmetry structure in the space of solutions. When solutions are 
written in terms of the Jacobi elliptic functions the symmetry is codified  in the functional form of its modulus, and is
described mathematically by the six dimensional coset  group $\Gamma/\Gamma(2)$ where $\Gamma$ is the 
modular group and $\Gamma(2)$ is its congruence subgroup of second level. In this paper we discuss   
the physical consequences this symmetry has on the pendulum motions and it is argued they have similar 
properties to the ones  termed as duality symmetries in other areas of physics, such as field theory and string theory. 
In particular a single solution of pure imaginary time for all allowed value of the total mechanical energy is given 
and obtained as the $S$-dual of a single solution of real time, where $S$ stands for the $S$ generator of the 
modular group. 
\end{abstract}

\maketitle

\section{Introduction}

The simple plane pendulum constitutes an important physical system whose analytical solutions are well known. 
Historically the first systematic study of the pendulum is attributed to Galileo Galilei, around 1602. Thirty years later 
he discovered that the period of small oscillations is approximately independent of the amplitude of the swing, 
property termed as isochronism, and in 1673 Huygens published the mathematical formula for this period. 
However, as soon as 1636, Marin Mersenne and Ren\'{e} Descartes had stablished that the period in fact does 
depend of the amplitude \cite{Matthews}. The mathematical theory to evaluate this period took longer to be 
established.
 
The Newton second law for the pendulum leads to a nonlinear differential equation of second order whose 
solutions are given in terms of either {\it Jacobi elliptic functions} or {\it Weierstrass elliptic functions}  
\cite{Whittaker1927,DuVal,Lang,Lawden,McKean,Armitage}. There are several textbooks on classical mechanics 
\cite{Appell2,Helmholtz,Whittaker:1917},  and recent papers \cite{Belendez,Brizard,Ochs}, that give account of 
these solutions. From the mathematical point of view the subject of interest is the one of  {\it elliptic curves} 
such as $y^2=(1-x^2)(1-k^2x^2)$, with $k^2 \neq 0,1$, the corresponding {\it elliptic integrals} 
$\int_0^x dx/y$ and the {\it elliptic functions} which derive from the inversion of them. Generically the 
domain of the elliptic functions is the complex plane $\mathbb{C}$ and they depend also on the value 
of the modulus $k$. The theory began to be studied in the mid eighteenth century and involved great 
mathematicians such as Fagnano, Euler, Gauss and Lagrange. The cornerstone in its development is due to 
Abel \cite{Abel} and Jacobi \cite{Jacobi,JacobiFundamenta}, who replaced the elliptic integrals by the elliptic 
functions as the object of study. Since then they both are recognized jointly as the mathematicians that developed 
the elliptic functions theory in their current form and to the theory itself as one of the jewels of nineteen-century 
mathematics. 

Because the solutions to the simple pendulum problem are given in terms of elliptic functions and the founder 
fathers of the subject taught us all the interesting properties of these functions, it can be concluded that all the 
characteristics of the different type of motions of the pendulum are known. This is strictly true, however most of 
the references on elliptic functions (see for instance \cite{Whittaker1927,DuVal,Lang,Lawden,McKean,Armitage} 
and references therein) focus, as it should be, on its mathematical properties, applying just some of them to the 
simple pendulum as an example. In this paper we review part of the analysis made by Klein \cite{Klein}, who 
studied  the properties that the transformations of the {\it modular group} $\Gamma$ and its 
{\it congruence subgroups} of finite index $\Gamma(N)$ have on the {\it modular parameter} $\tau$, being the 
latter a function of the quarter periods  $K$ and $K_c$ which in turn are determined by the value of the square 
modulus $k^2$. Our main interest in this paper is accentuate the physical meaning that these transformations 
have in the specific case of the simple pendulum, in our opinion this is a piece of analysis missing in the literature. 

For our purposes the relevant mathematical result is that the congruence subgroup of level 2, denoted as 
$\Gamma(2)$, is of order six in $\Gamma$ and therefore a fundamental cell for $\Gamma(2)$ can be formed 
from six copies of any fundamental region ${\cal F}$ of $\Gamma$ produced by the action of the six elements 
on the set of modular parameters $\tau$ that belong to ${\cal F}$. Each of these copies is distinguished from 
each other, according to the functional form of the modulus $k^2$ the six transformations leave invariant, being 
they: $k^2$, $1-k^2$, $1/k^2$, $1-1/k^2$, $1/(1-k^2)$ and $k^2/(k^2-1)$, interestingly these kind of relations 
appear in other areas of physics under the concept of {\it duality} transformations, nomenclature we will use here. 
This result can be understood from different mathematical points of view and provides a link between concepts  
such as lattices, complex structures on the topological torus ${\mathbb T}^2$, the modular group $\Gamma$ and 
elliptic functions. In the appendices we review briefly the basics of these concepts in order to keep the paper self 
contained as possible, emphasizing in every moment its role in the solutions of the simple pendulum. From the 
physical point of view, the pendulum can follow basically two kind of motions (with the addition of some limit 
situations), the specific type of motion depends entirely on the value of the total mechanical energy $k^2_E$, if 
$0<k^2_E<1$ the motion is oscillatory and if $1<k^2_E<\infty$ the motion is circulatory. Therefore in the problem 
of the simple pendulum, there are two relevant parameters, the square modulus $k^2$ of the elliptic functions that 
parameterize its solutions in terms of the time variable, and the total mechanical energy of the motion $k^2_E$. As 
we will discuss throughout the paper, the relation between these two parameters is not one-to-one due to the duality 
relations between the different invariant functional forms of $k^2$. For instance, for an oscillatory motion whose 
energy is $0<k^2_E<1$, it is possible to express the solution in terms of an elliptic Jacobi function whose square 
modulus is $k^2$, $1-k^2$, $1/k^2$, etc., in other words, the duality symmetries between the functional 
forms of the square modulus $k^2$ induce different equivalent ways to write the solution for an specific physical 
motion of the pendulum. The nature of the time variable also plays an important role in the equivalence of 
solutions, it turns out that whereas some solutions are functions of a real time, others are functions of 
a pure imaginary time. In this paper we will discuss all these issues and we will write down explicitly 
several equivalent solutions to describe an specific pendulum motion. These results constitute an example  in 
classical mechanics of a broader concept in physics termed under the name dualities. It is worth mentioning that 
many of the results we present here are already scattered throughout the mathematical literature but our exposition 
collects them together and is driven by a golden rule in physics that demands to explore all the physical 
consequences from symmetries. Notwithstanding some formulas have been worked out specifically for building 
up the arguments given in here and  to the best knowledge of the author they are not present in the available 
literature. As an example, we obtain a single solution that describes the motions of the simple pendulum as function 
of a pure imaginary time parameter, and we show it can be obtained through an $S$-duality transformation from a 
single formula that describes the motions of the simple pendulum for all permissible value of the total energy and 
which is function of a real time variable..

In a general context the duality symmetries we refer to, involve the special linear group $SL(2,\mathbb{Z})$ 
and appear often in physics either as an invariance of a theory or as a relationship among two different theories. 
Typically these discrete symmetries relate strong coupled degrees of freedom to weakly coupled ones and 
vice versa, and the  relationship is useful when one of the two systems so related can be analyzed, permitting 
conclusions to be drawn for its dual by acting with the duality transformations. There is a plethora of examples in 
physics that obey duality symmetries, which have led to important developments in field theory, gravity, statistical 
mechanics, string theory etc. (for an explicit account of examples see for instance \cite{Petropoulos:2012ne} and 
references therein). As a manner of illustration let us mention just two examples of theories that own duality 
symmetries:  i) in string theory appear three types of dualities, and the one that have the properties 
described above goes by the name $S$-duality, being the $S$ group element, one of the two generators of the 
group  $SL(2,\mathbb{Z})$ \cite{Schwarz:1996bh}. In this case the modular parameter $\tau$ is given by the coupling 
constant and therefore the $S$-duality relates the strong coupling regime of a given string theory to the weak 
coupling one of either the same string theory or another string theory. It is conjectured for instance that the 
type I superstring is $S$-dual to the $SO(32)$ heterotic superstring, and that the type $IIB$ superstring is  
$S$-dual to itself. ii) In 2D systems there is a broad class of dual relationships for which the electromagnetic 
response is governed by particles and vortices whose properties are similar. In particular for systems having 
fermions as the particles (or those related to fermions by the duality) the vortex-particle duality implies the duality 
group is the level-two subgroup $\Gamma_0(2)$ of $PSL(2,\mathbb{Z})$ \cite{Burgess:2000kj}.
The so often appearance of these duality symmetries in physics is our main motivation to heighten the fact that 
in classical mechanics there are also systems like the simple pendulum whose motions are related by 
duality symmetries. 

The structure of the paper is as follows. In section \ref{Pendulum} we summarize the real time solutions of the 
simple pendulum system in terms of elliptical Jacobi functions.  The relations between solutions with real time 
and pure imaginary time in terms of the $S$ group element of the modular group $\Gamma$ are exemplified in 
section \ref{Sduality} and the whole web of dualities is discussed in section \ref{Web}.  We make some final remarks 
in \ref{conclusions}. There are two appendices, appendix \ref{Modular} is dedicated to define the modular group, 
its congruence subgroups and its relation to double lattices whereas in appendix \ref{Jacobi} we give some 
properties of the elliptic Jacobi functions that are relevant for the analysis of the solutions of the simple pendulum.


\section{Real time solutions}\label{Pendulum}

The Lagrangian for a pendulum of point mass $m$ and length $l$, in a constant downwards gravitational field, of 
magnitude $-g$ ($g>0$), is given by
\begin{equation}
L(\theta,\dot{\theta})= \frac{1}{2}ml^2 \dot{\theta}^2-mgl(1-\cos \theta), 
\end{equation}
where $\theta$ is the polar angle measured counterclockwise respect to the vertical line and $\dot{\theta}$ stands 
for the time derivative of this angular position. Here the zero of the potential energy is set at the lowest vertical 
position of the pendulum, for which $\theta =2 n \pi$, with $n \in \mathbb{Z}$. 
The equation of motion for this system is 
\begin{equation}\label{eqmotion}
\ddot{\theta}+\frac{g}{l} \sin \theta = 0.
\end{equation}
This equation can be integrated once giving origin to a first order differential equation, whose physical 
meaning is the conservation of energy
\begin{equation}
E=\frac12 m l^2 \dot{\theta}^2+2mgl\sin^2 \left( \frac{\theta}{2} \right)=constant.
\end{equation}
Physical solutions exist only if $E \geq 0$. We can rewrite this equation of conservation, in dimensionless 
form, in terms of the dimensionless energy parameter: $k_E^2 \equiv \frac{E}{2mgl}$, and the dimensionless 
real time variable: $x \equiv \sqrt{\frac{g}{l}} t \in \mathbb{R}$, obtaining 
\begin{equation}\label{DimensionlessE}
\frac{1}{4} \left( \frac{d \theta}{d x} \right)^2 + \sin^2 \left( \frac{\theta}{2}\right) = k_E^2.
\end{equation}

Analyzing the potential, it is concluded that the pendulum has four different types of solutions depending of the 
value of the constant $k_E^2$. The analytical solutions in two of the four cases are given in terms of 
Jacobi elliptic functions and can be found for instance in 
\cite{Lawden, Armitage, Appell2,Helmholtz,Whittaker:1917,Belendez,Brizard,Ochs}. The other two 
cases can be considered just as limit situations of the previous two. The Jacobi elliptic functions 
are doubly periodic functions in the complex $z$-plane (see appendix \ref{Jacobi} for a short summary of 
the basic properties of these functions), for example, the function 
$\mbox{sn}(z,k)$ of square modulus $0 < k^2 < 1$, has the real primitive period $4K$ and the pure imaginary 
primitive period $2iK_c$,  where the so called quarter periods $K$ and $K_c$ are defined by the equations 
(\ref{DefK}) and (\ref{DefKc}) respectively. The properties of the different solutions are as follows:\\

$\bullet$ {\it Static equilibrium} ($\dot{\theta}=0$): The trivial behavior occurs when either $k_E^2=0$ or 
$k_E^2=1$. In the first case, necessarily $\dot{\theta}=0$. For the case $k_E^2=1$ we consider also the 
situation where $\dot{\theta}=0$. In both cases, the pendulum does not move, it is in static equilibrium. When 
$\theta =2 n \pi$ the equilibrium is stable and when  $\theta = (2n+1) \pi$ the equilibrium is unstable. \\

$\bullet$ {\it  Oscillatory motions} ($ 0< k_E^2<1$): In these cases the pendulum swings to and fro 
respect to a point of stable equilibrium. The analytical solutions are given by
\begin{eqnarray}
\theta (x)&=&2  \arcsin [k_E \,  \mbox{sn}(x-x_0,k_E)], \label{OscSol} \\
\frac{d \theta}{dx} \equiv \omega(x) &=& 2 \, k_E \,  \mbox{cn}(x-x_0,k_E), \label{OscP}
\end{eqnarray} 
where the square modulus $k^2$ of the elliptic functions is given directly by the energy 
parameter: $k^2 \equiv k^2_E$. 
Here $x_0$ is a second constant of integration and appears when equation 
(\ref{DimensionlessE}) is integrated out. It means physically that we can choose the zero of time arbitrarily. 
Derivatives of the basic Jacobi elliptic functions are given in (\ref{BasDer}). 

Without loose of generality, in our discussion we consider that the lowest vertical point of the oscillation 
corresponds to the angular value $\theta=0$, 
and therefore that $\theta$ takes values in the closed interval $[-\theta_m, \theta_m]$, where $0<\theta_m < \pi$ 
is the angle for which $\dot{\theta}_m=0$. This means that: 
$\sin(\theta/2) \in [-\sin(\theta_m/2), \sin(\theta_m/2)]$, where 
according to equation (\ref{DimensionlessE}),  $ \sin^2(\theta_m/2)= k^2_E <1$. Now according to (\ref{OscSol}) 
the solution is obtained by mapping \footnote{this map can be interpreted as a canonical transformation
see e. g. \cite{Reichl,0143-0807-36-5-055040}}: $\sin(\theta/2) \rightarrow k_E \,  \mbox{sn}(x-x_0,k_E)$, where
$x-x_0 \in [-K,K]$, or equivalently: $\mbox{sn}(x-x_0,k_E) \in [-1,1]$. With this map we describe half of a 
period of oscillation. To describe the another half, without loose of generality, 
we can extend the mapping in such a way that for a complete period of oscillation, $x-x_0 \in [-K,3K]$. 
Because the Jacobi function cn$(x-x_0,k_E) \in [-1,1]$, the dimensionless angular velocity $\omega(x)$ is 
restricted to values in the interval $[-2k_E,2k_E]$. 

\begin{figure}[htb]
\center{\includegraphics[height=3cm,width=9cm]{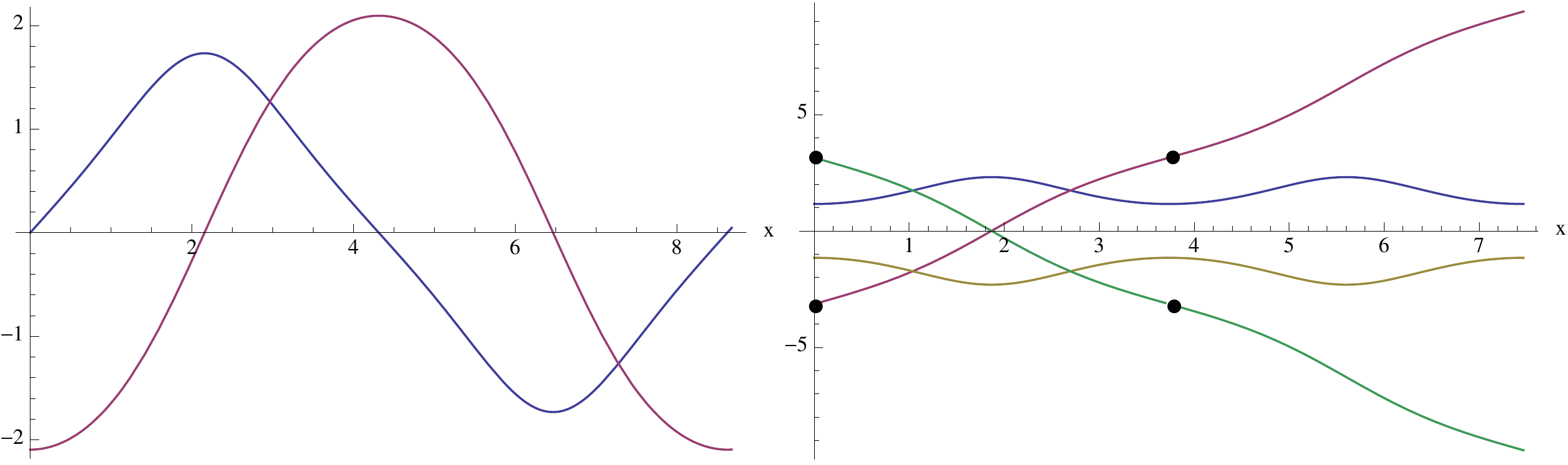}}
\caption{The first set of graphs represents an oscillatory motion of energy $k_E^2=3/4$ with 
$x_0=K \approx 2.1565$. $\theta(x)$ is given by the magenta graph and it oscillates in the interval 
$\theta(x)\in[-2\pi/3,2\pi/3]$. The angular velocity $\omega(x)$ is showed in blue and takes values in the 
interval $\omega(x) \in[-\sqrt{3},\sqrt{3}]$. Both graphs have period $4K$. The second set of graphs represents a 
couple of circulating motions both of energy $k_E^2=4/3$ with $x_0=K/k_E \approx 1.8676$.  
For the motion in the counterclockwise direction, the monotonic increasing function $\theta(x)$ 
is plotted in magenta, for the time interval $x \in [0,2x_0)$ it takes values in the interval  $\theta(x) \in [-\pi, \pi)$, 
whereas for the interval $x \in [2x_0,4x_0)$ it takes values in the interval $\theta(x) \in [\pi, 3\pi)$.
The angular velocity is showed in blue, it has period $2K/k_E$, is always positive and takes values in the interval 
$\omega(x) \in [2/\sqrt{3}, 4/\sqrt{3}]$. The other two plots represent a similar motion in the clockwise direction.} 
\label{GraphExam}
\end{figure}

As an example we can choose $x_0=K$, so at the time $x=0$, the pendulum is at minimum angular position 
$\theta(0)=-\theta_m$, with angular velocity $\omega(0)=0$. The pendulum starts moving from left to right, so at 
$x=K$ it reaches the lowest vertical position $\theta(K)=0$ at highest velocity $\omega(K)=2k_E$ and at $x = 2K$ 
it is at maximum angular position $\theta(2K)=\theta_m$ with velocity $\omega(2K)=0$. At this very moment the 
pendulum starts moving from right to left, so at $x= 3K$ it is again at $\theta(3K)=0$ but now with lowest 
angular velocity $\omega(3K)=-2k_E$ and it completes an oscillation at $x=4K$ when the pendulum reaches 
again the point $\theta(4K) =- \theta_m$ with zero velocity (see figure \ref{GraphExam}).  We can repeat this 
process every time the
pendulum swings in the interval $[-\theta_m, \theta_m]$, in such a way that the argument of the elliptic function 
$\mbox{sn}(x,k_E)$, becomes defined in the whole real line $\mathbb{R}$. It is clear that the period of the 
movement is $4K$, or restoring the dimension of time, $4K\sqrt{g/l}$.

Of course the value of $x_0$ can be set arbitrarily and it is also possible to parameterize the solution in such a way 
that at zero time $x=0$, the motion starts in the angle $\theta_m$ instead of $-\theta_m$. In this case the mapping of 
a complete period of oscillation can be defined for instance in the interval $x-x_0 \in [K,5K]$ and the initial condition 
can be taken as $x_0=-K$. In the discussion of the following section we will set $x_0=0$, so at time $x=0$, the 
pendulum is at the lowest vertical position ($\theta(0)=0$) moving from left to right. \\

$\bullet$ {\it Asymptotical motion} ($k_E^2=1$ and $\dot{\theta}\neq0$): In this case the angle $\theta$ 
takes values in the open interval  $(-\pi, \pi)$ and therefore, $\sin(\theta/2)  \in (-1,1)$. The particle just 
reach the highest point of the circle. The analytical solutions are given by
\begin{eqnarray}
\theta (x)&=& \pm  2  \arcsin [\tanh(x-x_0)], \label{AsinSol} \\
\omega(x) &=& \pm 2 \, \mbox{sech}(x-x_0).
\end{eqnarray} 
The sign $\pm$ corresponds to the movement from $(\mp \pi \rightarrow \pm \pi)$.  Notice that $\tanh(x-x_0)$, 
takes values in the open interval $ (-1,1)$ if: 
$x-x_0 \in (-\infty, \infty)$. For instance if $\theta \rightarrow \pi$, $x-x_0 \rightarrow \infty$ 
and $ \tanh(x-x_0)$ goes asymptotically to 1. It is clear that this movement is not periodic. In the 
literature it is common to take $x_0=0$.\\

$\bullet$ {\it Circulating motions} ($k_E^2>1$): In these cases the momentum of the particle is large 
enough to carry it over the highest point of the circle, so that it moves round and round the circle, always in 
the same direction. The solutions that describe these motions are of the form
\begin{eqnarray}
\theta (x)&=& \pm 2\,  \mbox{sgn}\left[\mbox{cn} \left( k_E (x-x_0) , \frac{1}{k_E} \right) \right] 
 \arcsin \left[ \mbox{sn} \left( k_E (x-x_0) , \frac{1}{k_E} \right) \right], 
\label{CirSol} \\
 \omega(x) &=& \pm 2 \, k_E \, \mbox{dn} \left( k_E (x-x_0) , \frac{1}{k_E} \right) \label{PCirSol},
\end{eqnarray} 
where the global sign $(+)$ is for the counterclockwise motion and the $(-)$ sign for the motion in the clockwise 
direction. The symbol sgn$(x)$ stands for the piecewise sign function which we define in the form
\begin{equation}
\mbox{sgn}\left[\mbox{cn} \left( k_E (x-x_0) , \frac{1}{k_E} \right) \right] = \left \{ 
\begin{array}{cc}
+1 & \mbox{if} \, \,  \, \, \, (4n-1)K \leq k_E (x-x_0) < (4n+1)K, \\
-1 & \mbox{if}  \, \,  \, \, \, (4n+1)K \leq k_E (x-x_0) < (4n+3)K,
\end{array}
\right.
\end{equation}
and its role is to shorten the period of the function sn$( k_E (x-x_0) , 1/k_E )$ by half, as we argue below. This fact 
is in agreement with the expression for the angular velocity $\omega(x)$ because the period of the elliptic function  
dn$( k_E (x-x_0) , 1/k_E )$ is $2K/k_E$ instead of $4K/k_E$ that is the period of the elliptic function  
sn$( k_E (x-x_0) , 1/k_E )$.

The square modulus $k^2$ of the elliptic functions is equal to the inverse of the energy parameter 
$0<k^2=1/k_E^2<1$. Without loosing generality we can assume both that $k_E(x-x_0)\in [-K,K)$, where $K$ 
is defined in (\ref{DefK}) and evaluated for $k^2=1/k^2_E$ and that  arcsin[sn$( k_E (x-x_0) , 1/k_E)] \in [-\pi/2,\pi/2)$. 
Because in this interval the function sgn[cn$( k_E (x-x_0) , 1/k_E)]=1$, the angular position function
$\theta(x) \in [\mp \pi, \pm \pi)$ for the global sign ($\pm$) in (\ref{CirSol}). As for the interval $k_E(x-x_0)\in [K,3K)$, 
we can consider that the function arcsin[sn$( k_E (x-x_0) , 1/k_E)] \in (-3\pi/2,-\pi/2]$, and because the function 
sgn[cn$( k_E (x-x_0) , 1/k_E)] =-1$, it reflects the angular position interval, obtaining finally that 
$\theta \in [\pm \pi, \pm 3\pi)$ for the global ($\pm$) sign in (\ref{CirSol}) (see figure \ref{GraphExam}).
We stress  that the consequence of flipping the sign of the angular interval through the sgn function is to make 
the function $\theta(x)$ piecewise periodic, whereas the consequence of taking a different angular interval for 
the image of the arcsin function every time its argument changes from an increasing to a decreasing 
function and vice versa, is to make the function $\theta(x)$ a continuous monotonic increasing (decreasing) 
function for the global sign + (-). Explicitly the angular position function changes as 
\begin{equation}
\theta\left( x+n \frac{2K}{k_E} \right)= \theta(x) \pm 2\pi n.
\end{equation}
It is interesting to notice that if we would not have changed the image of the arcsin function, the angular 
position function $\theta(x)$ would have resulted into a piecewise function both periodic and discontinuous. 

The angular velocity is a periodic function whose period is given by $T_{circulating}=2(K/k_E) \sqrt{g/l}$ which 
means as expected that higher the energy, shorter the period. Because the image of the Jacobi function 
dn$(x,k)\in [\sqrt{1-1/k_E^2},1]$, the angular velocity takes values in the interval 
$|\omega| \in [2\sqrt{k_E^2-1}, 2|k_E|]$.
An interesting property of the periods that follows from solutions (\ref{OscP}) and (\ref{PCirSol})
is that $T_{oscillatory} = k_E T_{circulating}$ where $k_E^2$ is the energy of a circulating motion 
and $k^2=1/k_E^2$ is the modulus used to compute $K$ in both cases. This is a clear hint that a relation 
between circulating and oscillatory solutions exists.

These are all the possible motions of the simple pendulum. It is straightforward to check that the solutions satisfy the 
equation of  conservation of energy (\ref{DimensionlessE}) by using the following relations between the Jacobi 
functions (in these relations the modulus satisfies $0<k^2 <1$) and its analogous relation for hyperbolic functions 
(which is obtained in the limit case $k=1$)
\begin{eqnarray}
\mbox{sn}^2(x,k) + \mbox{cn}^2(x,k) &=& 1, \label{relJac1}\\
\tanh^2(x)+\mbox{sech}^2(x) &=&1, \\
 k^2 \mbox{sn}^2(x,k)  + \mbox{dn}^2(x,k)  &=& 1. \label{relJac2}
\end{eqnarray}


\section{Imaginary time solutions and $S$-duality}\label{Sduality}

\begin{figure}[htb]
\center{\includegraphics[height=3cm,width=6cm]{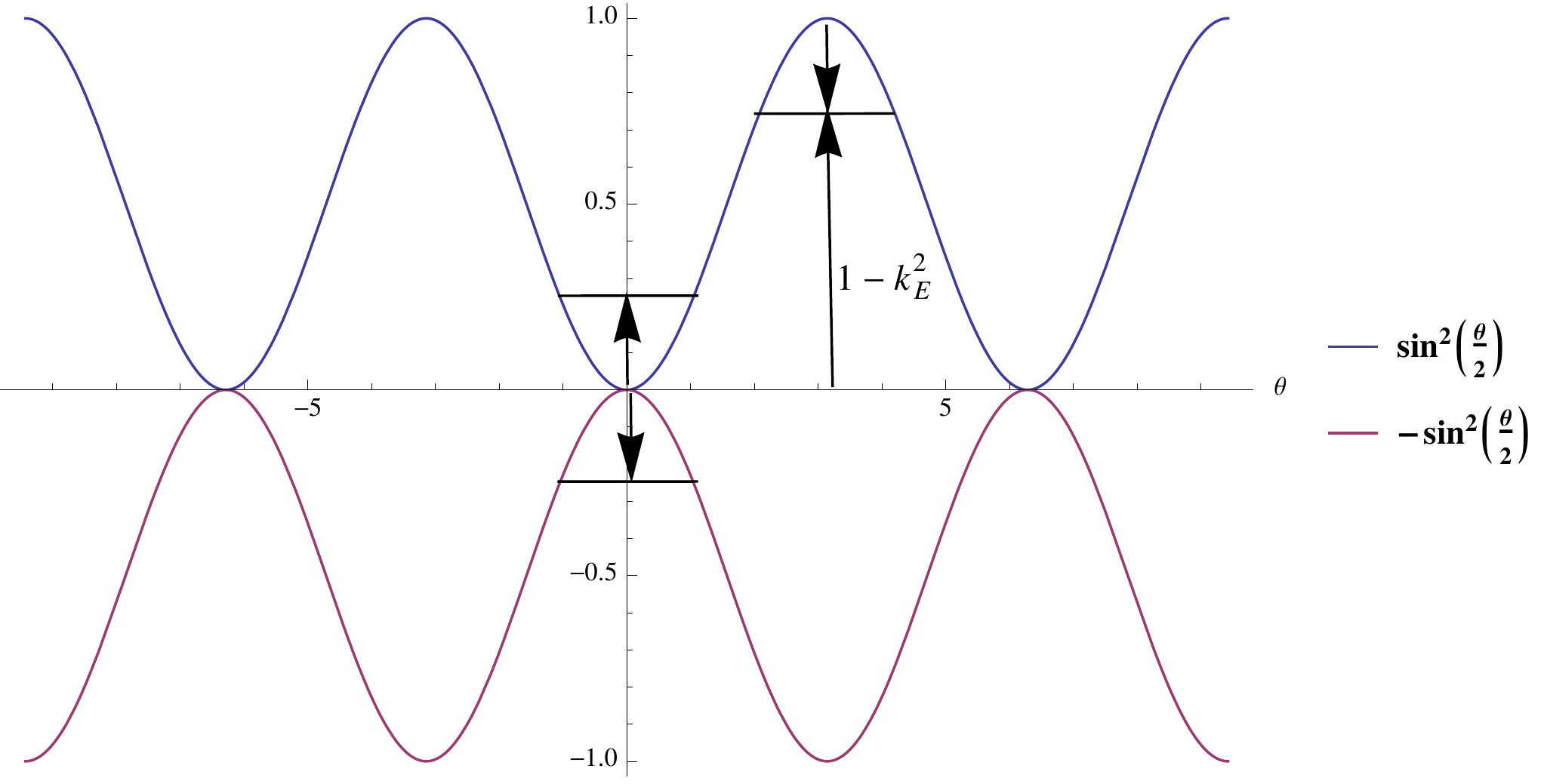}}
\caption{In the figure we show the pendulum potential (blue) for the dynamical motion parameterized with a real 
time variable. The inverted potential (magenta) corresponds to a dynamics parameterized by a pure imaginary time 
variable.} 
\label{InvPot}
\end{figure}

The argument $z$ of the Jacobi elliptic functions is defined in the whole complex plane $\mathbb{C}$ and the 
functions are doubly periodic  (see appendix \ref{Jacobi}), however in  the analysis above, time was considered 
as a real variable, and therefore in the solutions of the simple pendulum  only the real quarter 
period $K$ appeared. In 1878 Paul Appell clarified the physical meaning of the imaginary 
time and the imaginary period in the oscillatory solutions of the pendulum \cite{Appell,Armitage}, by introducing an 
ingenious trick, he reversed the direction of the gravitational field: $g \rightarrow -g$,  i.e. now the gravitational 
field is upwards. In order the Newton equations of motion remain invariant under this change in the force, we 
must replace the real time variable $t$ by a purely imaginary one: $\tau \equiv  \pm it$. Implementing these
changes in the equation of motion (\ref{eqmotion}) leads to the equation
\begin{equation}
\frac{d^2 \theta}{d\tau^2} - \frac{g}{l} \sin \theta =0.
\end{equation}
Writing this equation in dimensionless form requires the introduction of the pure imaginary time variable
$y \equiv  \pm \tau \sqrt{g/l}=\pm ix$. Integrating once the resulting dimensionless equation of motion gives origin to 
the following equation
\begin{equation}\label{ConsEner2}
\frac{1}{4} \left( \frac{d \theta}{dy} \right)^2 - \sin^2 \left( \frac{\theta}{2}\right) = E',
\end{equation}
which looks like equation (\ref{DimensionlessE}) but with an inverted potential (see figure \ref{InvPot}). 

We can solve the equation in two different but equivalent ways: i) the first option consists in writing down the equation 
in terms of a real time variable and then flipping  the sign of the whole equation in order to have positive energies, 
the resulting equation is of the same form as equation  (\ref{DimensionlessE}),  ii) the second option consists in 
solving the equation directly in terms of the imaginary time $y$.  Because both solutions describe to the same 
physical system, we can conclude that both are just different representations of the same physics. These two-ways 
of working provide relations between the elliptic functions with different argument and different modulus. As we 
will discuss the relations among different time variables and modulus can be termed as duality relations and 
because the mathematical group operation beneath these relations is the $S$ generator of the modular group 
$PSL(2,\mathbb{Z})$, we can refer to this duality relation as $S$-{\it duality}. 

\subsection{Real time variable}

If we write down (\ref{ConsEner2}) explicitly in terms of the real time parameter $x$, we obtain a conservation 
equation of the form
\begin{equation}\label{ConsEner3}
-\frac{1}{4} \left( \frac{d \theta}{dx} \right)^2 - \sin^2 \left( \frac{\theta}{2}\right) = E'.
\end{equation}
The first feature of this equation is that the constant $E'$ is negative $(E'<0)$. This happens because as a 
consequence of the imaginary nature of time, the momentum also becomes an imaginary quantity and when 
it is written in terms of a real time it produces a negative kinetic energy. On the other side the inversion of the force 
produces a potential modified by a global sign. Flipping the sign of the whole equation and denoting 
$E'=-k^2_E$, leads to the equation (\ref{DimensionlessE})
\begin{equation}\label{ConsEnerxReal}
\frac{1}{4} \left( \frac{d \theta}{dx} \right)^2 + \sin^2 \left( \frac{\theta}{2}\right) =k_E^2.
\end{equation}
We have already discussed the solutions to this equation (see section \ref{Pendulum}). However because 
we want to understand the symmetry between solutions, it is convenient to write down the ones of the 
circulating motions (\ref{CirSol})-(\ref{PCirSol})  relating the modulus of the Jacobi elliptic functions not to 
the inverse of the energy but to the energy itself, which can be accomplished by considering that 
the Jacobi elliptic functions can be defined for modulus greater than one. So we can write down both the 
oscillatory and the circulating motions in a single expression \cite{Ochs}
\begin{equation}\label{SolImReal}
\theta(x)= \pm2 k_E \, \mbox{sgn}[\mbox{dn}(x-x_0, k_E)]  \arcsin[\mbox{sn}(x-x_0, k_E)].
\end{equation}
Here the square modulus $k^2=k_E^2$ takes values in the intervals $0<k_E^2<1$ for the oscillatory motions and 
$1<k_E^2< \infty$ for the circulating ones. The reason of writing down the circulating solutions in this way is 
because introducing another group element of $PSL(2,\mathbb{Z})$, we can relate them to the standard form of 
the solution  (\ref{CirSol}) with modulus smaller than one. We shall do this explicitly in the next section.

\subsection{Imaginary time variable}

In order to solve  equation (\ref{ConsEner2}) directly in terms of a pure imaginary time variable, it is convenient 
to rewrite the equation in a form that looks similar to equation (\ref{ConsEnerxReal}), which we have already solved, 
and with this solution at hand go back to the original equation and obtain its solution. We start by shifting the value 
of the potential energy one unit such that its minimum value be zero. Adding a unit of energy to both sides of the 
equation leads to
\begin{equation}\label{ConsEneryIm}
\frac{1}{4} \left( \frac{d \theta}{dy} \right)^2 + \cos^2 \left( \frac{\theta}{2}\right) = 1-k_E^2.
\end{equation}
The second step is to rewrite the potential energy in such a form it coincides with the potential energy of 
(\ref{ConsEnerxReal}) and in this way allowing us to compare solutions. We can accomplish this by a simple 
translation of the graph, for instance by translating it an angle of $\pi/2$ to the right (see figure \ref{InvPot}). 
Defining $\theta'= \theta - \pi$, we obtain
\begin{equation}\label{ConsEner4}
\frac{1}{4} \left( \frac{d \theta'}{dy} \right)^2 + \sin^2 \left( \frac{\theta'}{2}\right) = 1-k_E^2.
\end{equation}
Solutions to this equation are given formally as
\begin{equation}\label{GenImSol}
\sin  \left( \frac{\theta'}{2}\right) = \pm \sqrt{1-k_E^2} \, \mbox{sn}\left( y-\tilde{y}_0, \sqrt{1-k_E^2} \right).
\end{equation}
Now it is straightforward to obtain the solution to the original equation (\ref{ConsEner2}), by going back to the 
original $\theta$ angle, obtaining  
\begin{equation}\label{SolImT}
\theta(x)=
\pm \, 2\, \mbox{sgn}\left[  \sqrt{1-k_E^2} \, \mbox{sn} \left( y-\tilde{y}_0, \sqrt{1-k_E^2} \right) \right]  
\arcsin \left[ \mbox{dn} \left( y-\tilde{y}_0, \sqrt{1-k_E^2} \right) \right].
\end{equation}
In this last expression we are assuming that equation (\ref{relJac2}) is valid for every allowed value of the energy 
$k^2_E \in (0,1) \cup (1,\infty)$, or equivalently  $1-k_E^2 \in (-\infty,0) \cup (0,1)$ (see equations (\ref{ksdmodim}) 
and (\ref{ksdmodinv})). It is important to stress that while $k_E^2$ has the interpretation of being an energy, 
$1-k_E^2$ can not be interpreted as such, as we will discuss below. Notice we have denoted to the 
integration constant in the variable $y$ as $\tilde{y}_0$ to emphasize that $\tilde{y}_0\in \mathbb{C}$ and is not 
necessarily a pure imaginary number. This happen because in contrast to the case of a real time variable where 
the integral along the real line $x \in \mathbb{R}$ can be performed directly, when the variable is complex it is 
necessary to chose a valid integration contour in order to deal with the poles of the Jacobi elliptic functions 
\cite{Whittaker1927}. For instance, the function dn$(y,k)$ has poles in $y=(2n+1)iK_c$ (mod $2K$) for 
$n \in \mathbb{Z}$, but dn$(ix+(2n+1)K , k)$ is oscillatory for every $x \in \mathbb{R}$ and $0<k<1$. The sign 
function in the solutions is introduced again in order to halve the period of the circulating motions respect to the 
oscillatory ones.

\subsection{Equivalent solutions}

In the following discussion we will assume without losing generality that $0<k^2 \leq 1/2$ and therefore that 
its complementary modulus is defined in the interval $1/2 \leq k_c^2<1$. The cases where $1/2 \leq k^2<1$ and 
therefore where $0<k_c^2 \leq 1/2$ can be obtained from the  case we are considering by interchanging the 
modulus and the complementary modulus $k^2 \leftrightarrow k_c^2$.\\

$\bullet$ {\it Oscillatory motion}: Let us consider oscillatory solutions for total mechanical energy 
$0<k_E^2= k^2 \leq 1/2$. Solutions for these motions can be expressed in terms of either i) a real time 
variable and given by equation (\ref{SolImReal}), or ii) in terms of a pure imaginary 
time variable. In the latter case the suitable constant is $\tilde y_0= iK-K_c$ and 
according to the equation (\ref{SolImT}) and due to the equivalence of solutions we have
\begin{equation}\label{SOsc1}
\theta_k(x) \equiv 2 \arcsin[ k \, \mbox{sn}(x, k)] =  2 \arcsin[ \mbox{dn}(ix-iK+K_c, k_c)] \equiv \theta_{k_c}(ix).
\end{equation}
This result is very interesting, it is telling us that any oscillatory solution can be represented as an elliptic  function 
either of a real time variable or a pure imaginary time variable and although they have the same energy, they differ 
in the value of its modulus. For solutions with real time the square modulus coincides with the energy 
$k_E^2$ and for solutions with pure imaginary time, the square modulus is equal to $1-k_E^2$. It 
is clear that the modulus of the two representations of an oscillatory solution satisfies the relation
\begin{equation}\label{ConsKyKc}
k^2 + k_c^2 = 1.
\end{equation}
As discussed in appendix \ref{Jacobi}, the elliptic function dn$(z,k_c)$ has an imaginary period $4iK$, therefore 
the period of the imaginary time oscillatory motion is $4iK \sqrt{g/l}$, which is in complete agreement with the 
periods  $4K \sqrt{g/l}$ for the solutions with real time. From equations (\ref{SOsc1}) it is straightforward to 
compute the angular velocity in terms of an elliptic function whose argument is a pure imaginary time variable 
(see table \ref{SSolutions}). A similar result is obtained for an oscillatory motion with energy $k_E^2=k_c^2$.

Two final comments are necessary, first in the general solutions (\ref{SolImReal}) and (\ref{SolImT}) the
$\pm$ signs appeared, however in (\ref{SOsc1}) there is not reference to them. This happen
because they are explicitly necessary only in the circulating motions. In the case of oscillatory motions the $(-)$ sign 
can be absorbed in the solution by rescaling the time variable in both cases (real and pure imaginary time). 
Regarding the elliptic function inside the sign function it does not appear because in the case of  (\ref{SolImReal}) we 
have sgn[dn$(x,k)]=1$ and also in (\ref{SolImT}) sgn[$k_c$ sn$(ix-iK_c+K, k)]=1$.\\

$\bullet$ {\it Circulating motion}: For the circulating motion we must also separate the energy ranks  
in two cases. If we are considering the solutions (\ref{SolImReal}) which have real time variable, the corresponding 
energy ranks are $1<k_E^2=1/k_c^2\leq 2$  and $2 \leq k_E^2=1/k^2 < \infty$. On the other side, if the solution
involves a pure imaginary time variable  (equation (\ref{GenImSol})) the relevant energies take values in the ranks 
$-1\leq 1-k_E^2=- k^2/k_c^2 <0$, and  $-\infty <1-k_E^2=- k^2_c / k^2 \leq -1$. Explicitly we have for the first rank
\begin{eqnarray}
\theta_{1/k_c}(x) & \equiv&  2 \, \mbox{sgn}\left [\mbox{dn} \left( x, \frac{1}{k_c} \right ) \right] 
\arcsin \left[  \frac{1}{k_c} \, \mbox{sn}\left( x, \frac{1}{k_c} \right ) \right] \nonumber \\
&=&  2  \, \mbox{sgn}\left [ \left( i \frac{k}{k_c} \right) \mbox{sn} \left( ix-iK,  i \frac{k}{k_c} \right) \right]  
\arcsin \left[  \mbox{dn} \left( ix-iK,  i \frac{k}{k_c} \right) \right] \equiv \theta_{ik/k_c}(ix).
\end{eqnarray} 
Notice that in a similar way to the oscillatory case, we have the following relations between the sum of the 
square modulus  
\begin{equation}
\frac{1}{k_c^2}-\frac{k^2}{k_c^2}=1.
\end{equation}
Analogous relations can be found for the solutions with energy $ k_E^2=1/k^2$ and for motions in the clockwise 
direction.

\subsection{$S$ group element as member of $PSL(2,\mathbb{Z})$}

It is possible to reach the same conclusions as in the previous subsection but this time following a slightly different 
path. In appendix \ref{Jacobi} we have summarized the action of the different group elements of 
$PSL(2,\mathbb{Z})$ on the Jacobi elliptic functions, in particular the action of the $S$ group element.
Starting for instance with a solution involving a real time variable and applying the action of the $S$ group element, 
it is possible to obtain the corresponding solution in terms of a pure imaginary time variable. As we will show, 
the obtained results coincide with the ones we have discussed. 

$\bullet$ {\it Oscillatory motion}:  In this case the starting point  is the solution (\ref{OscSol}) and its time derivative
(\ref{OscP}) which depends  on a real time variable and describe an oscillatory pendulum solution with energy 
$k^2_E$. To fix the discussion we choose $x_0=0$. Applying the Jacobi's imaginary transformations equations 
(\ref{TransJacS}) which are the transformations generated by the $S$ generator of the $PSL(2,\mathbb{Z})$ 
group, we obtain  
\begin{eqnarray}\label{ImOscSol}
k\,  \mbox{sn}(x,k) &=&-i k\,  \mbox{sc}(ix,k_c)= - k \, \mbox{nd}(i x+iK ,k_c) =
\mbox{dn}(ix-iK+K_c,k_c), \\
k \, \mbox{cn}(x,k) &=& k\,  \mbox{nc}(ix,k_c)=- i k\, k_c \, \mbox{sd}(i x+iK ,k_c) = 
-i k_c \, \mbox{cn}(ix-iK+K_c,k_c),
\end{eqnarray}
recovering relation (\ref{SOsc1}) with their respective expressions for its time derivative.
Notice that although the transformed functions have modulus $k_c$ they satisfy
\begin{equation}
\mbox{dn}^2(ix-iK+K_c,k_c)-k_c^2  \mbox{cn}^2(ix-iK+K_c,k_c) =  k^2,
\end{equation}
which is telling that the solution is indeed of oscillatory energy $k_E^2=k^2$ as it should be. An analogous result 
is obtained if we start instead with a solution of modulus $k_c^2$ and real time variable.\\

$\bullet$ {\it Circulating case}: In the circulating case we have a similar story, under an $S$ transformation the 
circulating solutions (\ref{CirSol})-(\ref{PCirSol}) lead to the set
\begin{eqnarray}\label{ImCirSol}
\mbox{sgn} \left[  \mbox{dn} \left(k x, \frac{1}{k} \right)  \right] \frac{1}{k} \mbox{sn} \left( k x, \frac{1}{k} \right) &=& 
\mbox{sgn} \left[  i \frac{k_c}{k}  \, \mbox{sn} \left( i x-iK , i \frac{k_c}{k}  \right) \right] 
\mbox{dn} \left( i x-iK  , i \frac{k_c}{k} \right), \\
\mbox{cn} \left(k x, \frac{1}{k} \right) &=& k_c \mbox{cn} \left( i x-iK , i \frac{k_c}{k}  \right),
\end{eqnarray} 
which coincide with the solutions (\ref{SolImT}) for a choice of the constant $\tilde y_0 =iK$.


\section{Web of dualities}\label{Web}

\subsection{The set of $S$-dual solutions}

We have argued that a symmetry of the equation of motion for the simple pendulum 
leads to the possibility that its solutions can be obtained in two ways: i) considering 
a real time variable and ii) considering a pure imaginary time variable. The solutions for energies in the 
rank $k_E^2 \in (0,1) \cup (1,\infty)$ are given by Jacobi elliptic functions, the ones for energies   $k_E^2 \in (0,1)$ 
describe oscillatory motions and the ones for energies  $k_E^2 \in (1,\infty)$ describe circulating ones. On the other 
hand we also know that the Jacobi elliptic functions are doubly periodic functions in the complex plane 
$\mathbb{C}$  (see appendix \ref{Jacobi}), and additionally  to the complex argument $z$, they also depend on the 
value of the modulus whose square $k^2$ takes values in the real line $\mathbb{R}$ with exception of the points 
$k^2\neq 0$ and 1. In the previous section we have discussed that given a type of motion, for instance an 
oscillatory motion with energy $0<k_E^2\leq 1/2$, there are at least two equivalent angular functions describing it, 
one with modulus $k=k_E$ and real time denoted as $\theta_k(x)$ in (\ref{SOsc1}) and a second one with 
modulus $k_c=\sqrt{1-k^2_E}$ and  pure imaginary time denoted as $\theta_{k_c}(ix)$. We can refer to this 
dual description of the same solution as $S$-duality. In table  \ref{SSolutions} we give the 
solutions for all the simple pendulum motions (oscillatory and circulating) in terms of real time and its $S$-dual 
solution given in terms of a pure imaginary time.
\begin{table}[htb] 
\begin{tabular}{|c | c | c | c | c |} 
\hline
Energy $k_E^2$ & Variable & Real time solution & Imaginary time solution  \\
\hline 
\hline
$k^2 \in (0,1/2]$ & $\theta/2$ & $\arcsin[k$ sn$(x,k)]$ & $\arcsin$[dn$(ix-iK+K_c,k_c)]$   \\
 & $\omega/2$ & $k$  cn$(x,k)$ & $-ik_c$ cn$(ix-iK+K_c,k_c)$  \\
$1-k^2 \in [1/2,1)$ & $\theta/2$  &$\arcsin[k_c$ sn$(x,k_c)]$ & $\arcsin[$dn$(ix-iK_c+K,k)]$  \\
 & $\omega/2$ & $k_c$ cn$(x,k_c)]$ & $-ik$ cn$(ix-iK_c+K,k)$ \\
$\frac{1}{1-k^2} \in (1,2]$ & $\theta/2$ & $\pm$ sgn[dn$(x,1/k_c)] \arcsin[$sn$(x,1/k_c)/k_c]$ &   
$\pm$ sgn[$(ik/k_c)$ sn$(ix-iK_c, i k/k_c )]  \arcsin[$\mbox{dn} $(ix-iK_c, i k/k_c )]$  \\
 & $\omega/2$ & $\pm (1/k_c)$ cn$(x,1/k_c)$ &  $ \pm (k/k_c)$  \mbox{cn} $(ix-iK_c, i k/k_c )$  \\
$\frac{1}{k^2} \in [2,\infty)$ & $\theta/2$ & $\pm$ sgn[dn$(x,1/k)] \arcsin[$sn$(x,1/k)/k]$ 
& $\pm$ sgn[$(ik_c/k)$ sn$(ix-iK, i k_c/k )$] $\mbox{dn} (ix-iK, i k_c/k )$ \\ 
 & $\omega/2$ & $\pm (1/k)$ cn$(x,1/k)$& $\pm (k_c/k)$  \mbox{cn} $(ix-iK, i k_c/k )$ \\
\hline
\end{tabular}
\caption{The third column shows the solutions to the simple pendulum problem in terms of a real time variable 
when the total mechanic energy of the motion and the square modulus of the Jacobi elliptic function are the 
same. The fourth column shows its $S$-dual solutions in terms of a pure imaginary time variable.}
\label{SSolutions}
\end{table}

The fact that the solutions involve either real time or pure imaginary time only, but not a general complex time 
leads to the conclusion that although the domain of the elliptic Jacobi functions are all the points in a fundamental 
cell,  or due to its doubly periodicity, in the full complex plane $\mathbb{C}$, the pendulum solutions 
take values only in a subset of this domain. Let us exemplify this fact for a vertical fundamental cell, i.e., 
for values of the square modulus in the interval $0<k^2<1/2$, which correspond to a normal lattice $L^*$ (see 
appendix \ref{Jacobi}).  In this case the generators are given by $4K$ and $4iK_c$ with $K_c>K$. If the time 
variable $x$ is real, the solutions are given by the function sn($x,k$) which owns a pure imaginary period 
$2iK_c$. The oscillatory solutions on the fundamental cell are given generically either by 
$\arcsin[k$ sn($x-x_0,k$)] or $\arcsin[k$ sn($x-x_0+2iK_c,k$)], or in general on the complex plane $\mathbb{C}$ 
the domain of these solutions is given by all the horizontal lines whose imaginary part is constant and given by 
$2niK_c$ with $n \in \mathbb{Z}$. 
According to table \ref{SSolutions}, the oscillatory solutions of pure imaginary time on the same fundamental cell, 
have energies in the interval $1/2 \leq k^2_E=k^2_c <1$ and are given generically by $\arcsin[$dn($ix-ix_0+K,k$) 
or $\arcsin[$dn($ix-ix_0+3K,k$). In general the domain of these solutions in the 
complex plane $\mathbb{C}$ are all the vertical lines whose real part is constant and given by $(2n+1)K$ with 
$n \in \mathbb{Z}$, which is in agreement with the fact that the function dn$(z,k)$ owns a real period $2K$. Any other 
point in the domain of the elliptic Jacobi functions, different 
to the ones mentioned  do not satisfy the initial conditions of the pendulum motions. This discussion can be 
extended to the horizontal fundamental cells (normal lattices $iL^*$) whose modulus is given by $k_c$ 
and the ones that involve an $STS$ transformation and therefore a Dehn  twist (see appendix \ref{Jacobi}). 

We conclude that if we consider only solutions of real time variable such that the square modulus and the energy 
coincide (the four types of table \ref{SSolutions}), then 
the corresponding domains are horizontal lines on the normal lattices $L^*$, $iL^*$, $kL^*$ and $ik_cL^*$. 
If instead we consider the four solutions of pure imaginary time parameter, the corresponding 
domains are vertical lines on the normal lattices $iL^*$, $L^*$, $ikL^*$ and $k_cL^*$. However 
due to the fact that the modular group relates the normal lattices one to each other, we can consider less normal 
lattices and instead consider other Jacobi functions on the smaller set of normal lattices to obtain the same four 
group of solutions. We shall address this issue below.

\subsection{The lattices domain}

\begin{figure}[htb] 
\center{\includegraphics[height=4cm,width=3.0cm]{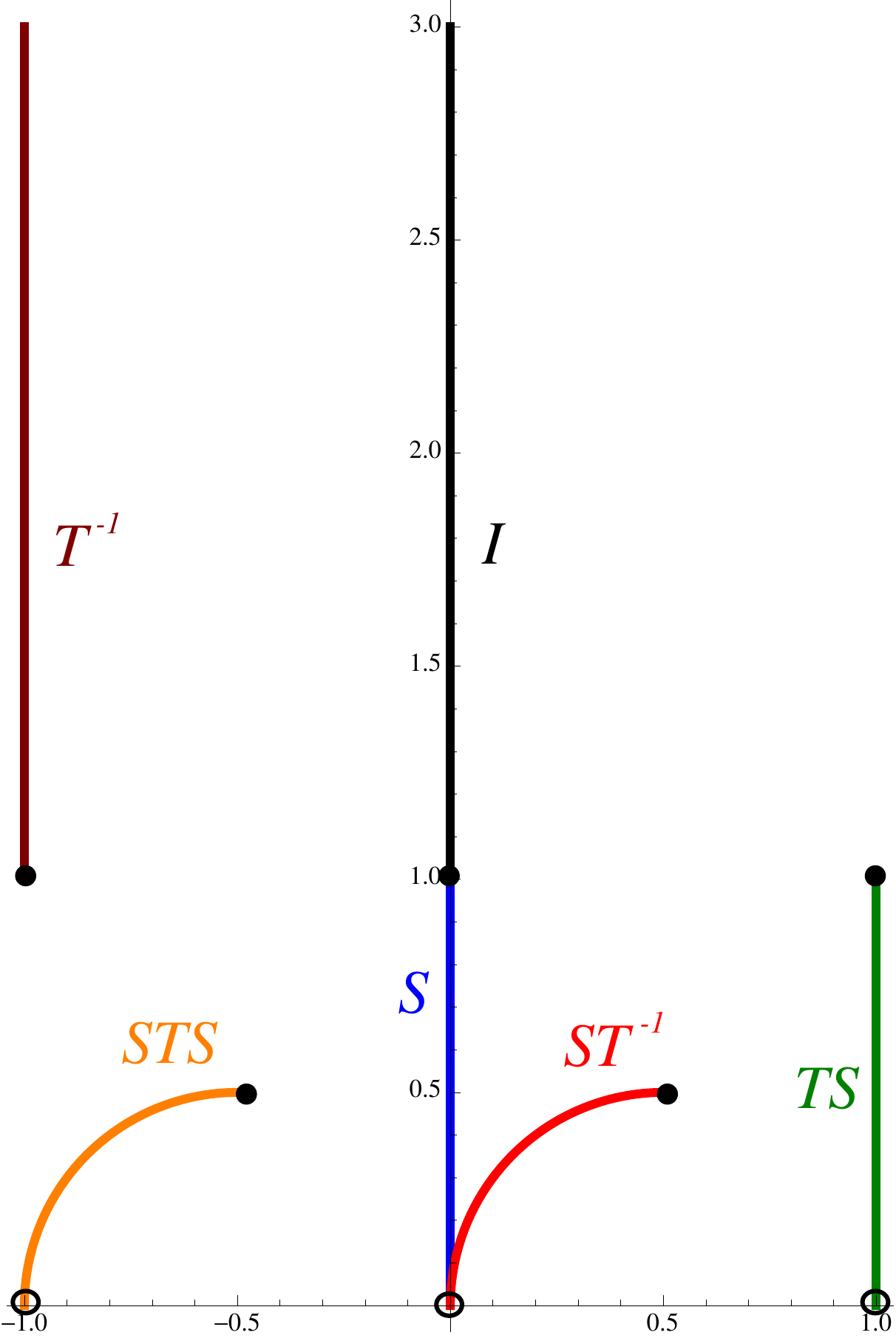}}
\caption{Figure shows the whole domain of values that the modular parameter $\tau$ can take for the Jacobi elliptic 
functions. This domain is a subset of the ${\cal F}_2$ fundamental region (figure \ref{RegionG2}). Black dots 
represent the values of the square modulus $k^2=1/2$, $1-k^2=1/2$, $1/k^2=2$, $1/(1-k^2)=2$, $1-1/k^2=-1$ and 
$k^2/(k^2-1)=-1$.} 
\label{LatCel}
\end{figure}

At this point it is convenient to discuss the domain of the lattices that play a role in the elliptic Jacobi functions and 
therefore in the solutions of the simple pendulum. As discussed in the appendix \ref{Modular}, the quarter 
periods of a Jacobi elliptic function whose square modulus is in the interval $0<k^2\leq1/2$, generate vertical 
lattices represented by a modular parameter of the form $\tau=iK_c/K$. The point $\tau=i$ is associated to the 
case where the rectangular lattice becomes square and corresponds to the value $k^2=1/2$. 
The set of all these lattices (black line in figure \ref{LatCel}) is represented in the complex plane by the left 
vertical boundary of the region ${\cal F}_1$ (figure \ref{RegionG2}) since the quotient $K_c/K \in [1,\infty)$. 
Acting on these values of the modular parameter with the six group elements of 
$PSL(2, \mathbb{Z}/2 \mathbb{Z})$ produce the whole set of values of the modular parameter (figure \ref{LatCel}) 
that are consistent with the elliptic Jacobi functions. For example, acting with the $S$ group element of $\Gamma$ 
on the vertical line  $\tau=iK_c/K$, generates the blue vertical line described mathematically by the set of modular 
parameters $\tau=iK/K_c$, with $K/K_c \in (0,1]$. It is clear that the set of six lines is a subset of the ${\cal F}_2$ 
fundamental region  and constitutes the whole lattice domain of the elliptic Jacobi functions.

In table \ref{Numerica} we give the numerical values (approximated) of the 
generators of the fundamental cell as well as the modular parameter for some values of the square modulus.

\begin{table}[htb]
\begin{tabular}{|c | c | c | c | c | c | c |} 
\hline
Modulus $k^2$ & $\omega_1/4$ & $\omega_2/4$ & $\tau$ \\
\hline 
\hline
0 & $\pi/2$ & $i \cdot \infty$ & $i \cdot \infty$ \\
1/4 & 1.68575 & $(2.15652) \, i$ & $1.27926 \, i$ \\ 
1/2 & 1.85407 & $(1.85407) \, i$& $i$  \\
3/4 & 2.15652 & $(1.68575) \, i$& $0.78170 \, i$  \\
1 & $\infty$ & $i\, \pi/2$ & 0  \\
4/3 & $2.87536+i \, 2.24767$ & $(2.24767)\, i$ & $ 0.37929+i \, 0.48521$ \\
2 & $3.70814(1\pm i)$ &  $(3.70814) \, i$ & $\pm 0.5+i\, 0.5$ \\ 
4 &$6.743-i \, 8.62608$ & $(8.62608) \, i$ & $-0.62071+i\, 0.48521$ \\
\hline
\end{tabular}
\caption{Approximated numerical values of the periods and the modular parameter for some real values of the 
modulus of the Jacobi elliptic functions. The values $k=0$ and $k=1$ correspond to limit situations where one of 
the two periods is lost. The value $k^2=1/2$ is known as a fixed point, it belongs both to the boundary of the 
regions ${\cal F}_1$ and $S$ of the fig. \ref{RegionG2} and is represented by the black dot  whose coordinates 
are $(0,i)$ in fig. \ref{LatCel}. The value $k^2=2$ is degenerated in the sense it can be 
represented  by two different types of fundamental cells, in one case the cell belongs to the boundary of the region 
$STS$ and in the another case it belongs to the boundary of $ST^{-1}$. The fundamental cell for some 
values of $k^2$ in this table are plotted in fig. \ref{CellS}.}\label{Numerica}
\end{table}

As a conclusion, for every value of the parameter $0<k^2\leq1/2$ there are six normal lattices related one to each 
other by transformations of the modular group. Therefore each solution of the simple pendulum with real time 
variable, showed in table \ref{SSolutions}, can be written in six different but equivalent ways, where each one of the 
six forms is in one to one correspondence with one of the six normal lattices. Their $S$-dual  solutions 
(see table \ref{SSolutions}) which are functions of a pure imaginary time are just one of the six different ways in 
which solutions can be written.

\subsection{$STS$-duality}

The form of the solutions for the simple pendulum expressed in table \ref{SSolutions} does not 
coincide with the expressions given in section \ref{Pendulum}, which by the way, are the standard form in 
which the solutions are commonly written in the literature.  In order to reproduce the standard form it is 
necessary to introduce the $STS$ transformation (see appendix \ref{Jacobi}). This transformation takes for 
instance a Jacobi function with modulus $0<k<1$ into a Jacobi function with modulus greater than one 
$1<1/k<\infty$. 
Taking the inverse transformation it is possible to take a Jacobi function with modulus $1<1/k$ into one with 
modulus $k<1$. Using the relations of the appendix \ref{Jacobi} it is straightforward to obtain equations 
(\ref{UsefulSTS}) which written in terms of $k_E$ instead of $k$  (remember than in this case $1<1/k=k_E$ 
$\Rightarrow$ $k=1/k_E<1$), lead to
\begin{equation}
k_E  \, \mbox{sn} \left(x,k_E \right) = \mbox{sn} \left( k_E \, x , 1/k_E \right), \hspace{0.5cm}
k_E  \, \mbox{cn} \left(x,k_E \right) = \mbox{dn} \left( k_E \, x , 1/k_E \right), \hspace{0.5cm}
\mbox{dn} \left(x,k_E \right) = \mbox{cn} \left( k_E \, x , 1/k_E \right).
\end{equation}
Inserting this relations in the circulating solutions of table \ref{SSolutions} reproduce solutions (\ref{CirSol}) and 
(\ref{PCirSol}). 

What we have done is to use the $STS$-duality between lattices and transform two of them 
$kL^*$ and $ik_cL^*$ into $L^*$ and $iL^*$. Restricted to solutions with real time, two of the four type of solutions 
for which $k^2=k_E^2>1$, are transformed to solutions for which $k^2=1/k_E^2<1$. As we have discussed  the 
domain of the solutions with real time variable are horizontal lines in the normal lattices $L^*$ and $iL^*$, thus in 
order to keep the four different types of solutions it is necessary to evaluate two different set of Jacobi 
functions (\ref{OscSol}) and (\ref{CirSol}) on the domain of each one of the two normal lattices $L^*$ and $iL^*$. 
It is clear that this is not the only way we can proceed, in fact we can transform the oscillatory solutions with $k<1$ 
into oscillatory solutions with modulus grater than 1. A similar analysis follows if we consider only solutions with 
imaginary time.

\subsection{A single normal lattice}

It is natural to wonder about the minimum number of normal lattices needed to express all the solutions of the 
simple pendulum. Due to the duality symmetries between lattices this number is one. As an example, 
if we now use the $S$-duality to relate the normal horizontal lattice $iL^*$ to the normal vertical lattice $L^*$,  
the horizontal lines that compose the domain in the horizontal lattice becomes vertical lines in the vertical lattices, 
which means to consider solutions with imaginary time in $L^*$. Thus we can end up with only one normal lattice 
and in order  to have the four different types of solutions, it is necessary to consider the whole domain of the lattice, 
i.e. both vertical lines (imaginary time) and horizontal lines (real time) and on each set of lines to consider two 
different solutions one oscillatory and one circulating. For completeness in table \ref{SolOneLattice} we give 
the four type of solutions in terms of only one value of the modulus
\begin{table}[htb] 
\begin{tabular}{|c | c |} 
\hline
Energy interval & Solution $\theta$  \\
\hline 
\hline
$k_E^2 \in (0,1/2]$ & 2 $\arcsin[k$ sn$(x,k)]$ \\
$k_E^2 \in [1/2,1)$ & 2 $\arcsin[$dn$(ix-iK_c+K,k)]$  \\
$k_E^2 \in (1,2]$ & $\pm$ 2 sgn[$(-ik/k_c)$ cn$(ix/k_c-iK_c/k_c, k)]  \arcsin[(1/k_c)$ \mbox{dn}$(ix/k_c-iK_c/k_c, k)]$\\
$k_E^2 \in [2,\infty)$ & $\pm$ 2 sgn[cn$(x/k, k )] \arcsin$[sn$(x/k, k )]$ \\ 
\hline
\end{tabular}
\caption{Solutions to the simple pendulum problem written in a unique lattice of square modulus 
$0<k^2\leq 1/2$.}\label{SolOneLattice}
\end{table}

It is clear that we can express all the solutions also for the other five different functional forms of the square modulus.


\section{Final remarks}\label{conclusions}

In this paper we have addressed the meaning of the fact that the complex domain of the solutions of the 
simple pendulum is not unique and in fact they are related by the $PSL(2,\mathbb{Z}/2\mathbb{Z})$ group, finding 
that the important issue for express the solutions is the relation between the values of the square modulus $k^2$ 
of the Jacobi elliptic functions, and the value of the total mechanical energy $k^2_E$ of the motion of the pendulum. 
Due to the symmetry  we conclude that there are six different expressions of the square modulus that are related 
one to each other trough the six group elements of $PSL(2,\mathbb{Z}/2\mathbb{Z})$. These six group actions can 
be termed as duality-transformations and therefore we have six dual representations of $k^2$. As a consequence  
there are six different but equivalent ways in which we can write an specific pendulum solution, 
and abusing a little bit of the language we could say there are duality relations between solutions.
This analysis teach us the lesson that we can restrict the domain of lattices to the ones whose 
modular parameter is in the pure imaginary interval $\tau \in i (1,\infty)$, or equivalently that we can 
express every solution of the simple pendulum either oscillatory or circulating with Jacobi elliptic functions 
whose value of the square modulus is in the interval $0<k^2 \leq 1/2$ (see table \ref{SolOneLattice}).

It is well known that there are several physical systems in different areas of physics whose solutions are also given 
by elliptic functions, for instance in classical mechanics some examples are the spherical pendulum, the Duffing 
oscillator, etc., in Field Theory the Korteweg de Vries equation, the Ising model, etc., 
\cite{Brizard,Petropoulos:2012ne}. It would be very interesting to investigate on similar grounds to the ones 
followed here, the physical meaning of the symmetries of the elliptic functions in these systems.


\appendix

\section{The modular group and its congruence subgroups} \label{Modular}

\subsection{The modular group}

The modular group $\Gamma$ is the group defined by the linear fractional transformations on 
the {\it modular parameter} $\tau \in \mathbb{C}$ (see for instance 
\cite{DuVal,Lang,Serre,Lawden,McKean,Armitage,Diamond} and references therein)
\begin{equation}\label{ModGroup}
\tau  \mapsto \Gamma(\tau)=\frac{a \tau +b}{c \tau +d},
\end{equation}
where $a,b,c,d \in \mathbb{Z}$ satisfying $ad-bc=1$, and the group operation is function 
composition. These maps all transform the real axis of the $\tau$ plane (including the point at infinity) into itself, 
and rational values into rational values. The group has two generators defined by the transformations
\begin{equation}\label{Generators}
S(\tau) \equiv - 1/\tau, \hspace{0.5cm}  \mbox{and}  \hspace{0.5cm}  T(\tau) \equiv 1 + \tau.
\end{equation}
The modular group is isomorphic to the projective special linear group $PSL(2, \mathbb{Z})$, which is 
the quotient of the 2-dimensional special linear group $SL(2, \mathbb{Z})$ by its center 
$\{ \mathbb{I},-\mathbb{I}\}$. In other words, $PSL(2, \mathbb{Z})=SL(2, \mathbb{Z})/\mathbb{Z}_2$ consists of 
all matrices of the form 
\begin{equation}\label{GenMatrix}
A=\left( \begin{matrix}
a & b \\
c & d 
\end{matrix} \right),
\end{equation} 
with unit determinant, and pair of matrices $A$, $-A$, are considered to be identical. The group 
operation is multiplication of matrices and the generators accordingly with (\ref{Generators}) are
\begin{equation}
S=\left( \begin{matrix}
0 & -1 \\
1 & \, \, \, \, 0 
\end{matrix} \right), 
\hspace{1cm}
T=\left( \begin{matrix}
1 & 1 \\
0 & 1 
\end{matrix} \right) .
\end{equation}
These group elements satisfy $S^2=(ST)^3=-\mathbb{I} \sim \mathbb{I}$ and $T^n=\left( \begin{matrix}
1 & n \\
0 & 1 
\end{matrix} \right)$.

\begin{figure}[htb] 
\center{\includegraphics[height=4cm,width=6.0cm]{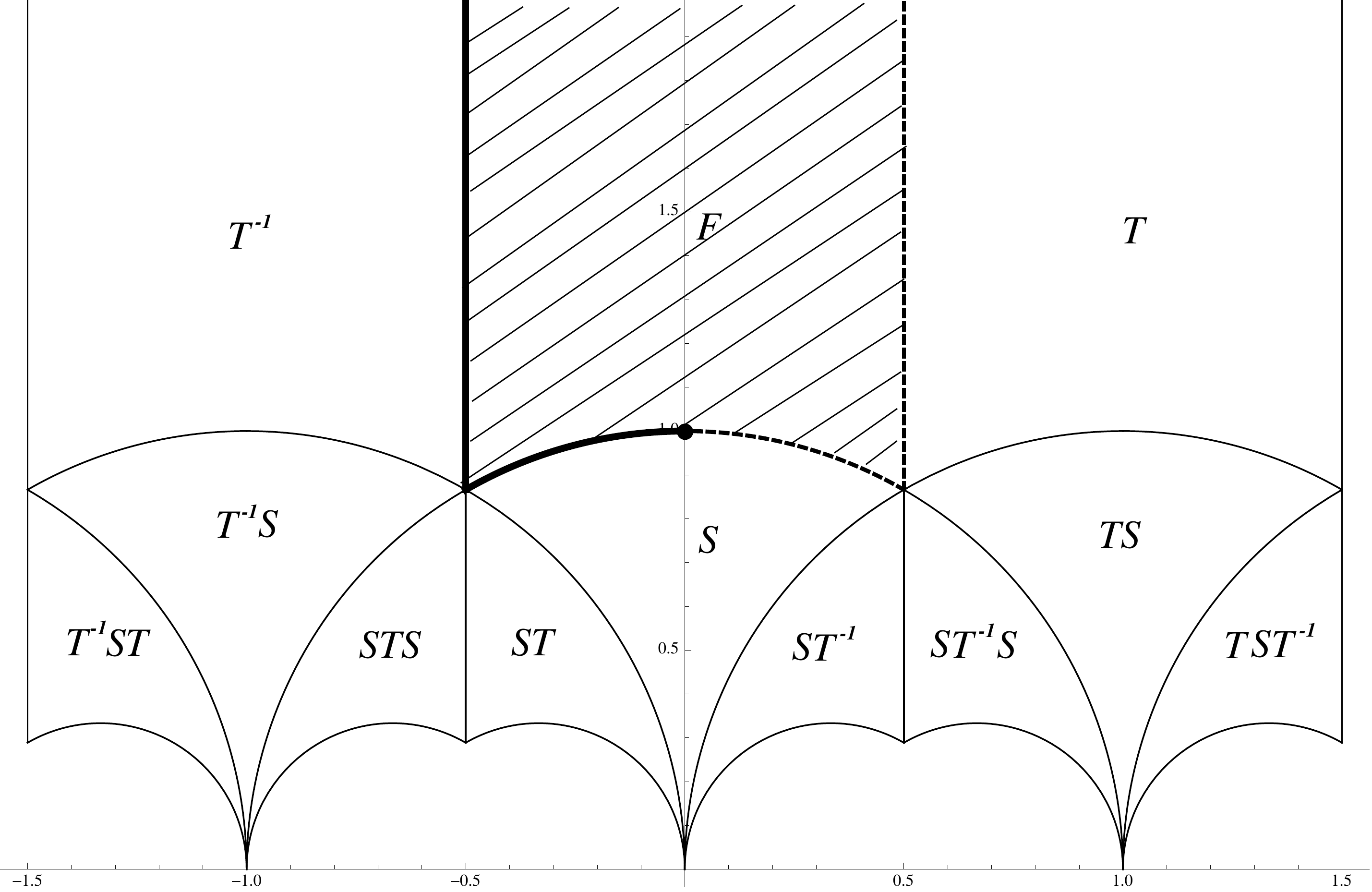}}
\caption{Tessellation of ${\cal H}$. The fundamental region ${\cal F}$ is represented by the shaded area and 
the heavy part of the boundary. This region is mapped to the whole upper plane $\mathbb{C}$ by the modular 
group $\Gamma$. The region can be viewed as a complete list of the inequivalent complex structures on the 
topological torus since conformal equivalence of tori is determined  by the modular equivalence of their period 
ratios. In the figure we show some copies of the fundamental region obtained by application of some group elements 
of $PSL(2, \mathbb{Z})$.}  
\label{FundamentalDomain}
\end{figure}

One important property of the modular group is that the upper half plane of $\mathbb{C}$, usually 
denoted as $\cal{H}$ and defined as ${\cal H} \equiv \{ z \in \mathbb{C} : Im(z) > 0\}$, can be generated by the 
elements of $PSL(2, \mathbb{Z})$  from a {\it fundamental domain or region} ${\cal F}$. Mathematically this
region is the quotient space ${\cal F}={\cal H}/PSL(2, \mathbb{Z})$ and satisfies two properties: (i) ${\cal F}$ is 
a connected open subset of ${\cal H}$ such that no two points in ${\cal F}$  are related by a $\Gamma$ 
transformation (\ref{ModGroup}) and (ii) for every point in ${\cal H}$ there is a group element $g \in \Gamma$  
such that  $g \tau \in {\cal F}$. There are many ways of constructing ${\cal F}$,  and the most common one found in 
the literature is to take the set of all points $z$ in the open region $\{ z: -1/2 < \mbox{Re}(z) < 1/2 \cap |z| >1 \}$,
union ``half'' of its boundary, for instance, the one that includes the points: $z=-1/2 + iy$ with 
$y \geq \sin(2\pi/3)$, and $|z|=1$ with $-1/2 \leq $ Re$(z) \leq 0$ (see figure \ref{FundamentalDomain}).
It is assumed that the imaginary infinite is also included.  

Geometrically, $T$ represents a shift of ${\cal F}$ to the right by 1, while $S$ represents the inversion of ${\cal F}$ 
about the unit circle followed by reflection about the imaginary axis. As an example, the figure 
\ref{FundamentalDomain} represents the transformations of the fundamental region ${\cal F}$ by the elements of 
the group: $\{ \mathbb{I}, T, T^{-1}, S,TS,T^{-1}S, ST, ST^{-1}, ST^{-1}S, TST^{-1},STS, T^{-1}ST \}$ 
\cite{Serre}.    
Notice that these 12 elements are all the independent ones that we can construct as iterative products of 
$S$, $T$ and $T^{-1}$ without powers of any of them involved  ($S^{-1}$ is simply $-S \sim S$ and therefore 
is not a different modular transformation). 
The other two transformations we can construct are not independent $TST=-ST^{-1}S$ and 
$T^{-1}ST^{-1}=STS$. Further products of the generators with these transformations give us the whole 
tessellation of the upper complex plane. In particular the orbit of the points Im$(z) \rightarrow \infty$ are 
the rational numbers $\mathbb{Q}$ and are called {\it cusps}.

\subsection{Congruence subgroups}

Relevant for our discussion are the {\it congruence subgroups of level} $N$ denoted as $\Gamma(N)$ 
(or $\Gamma_N$).  They are defined as subgroups of the modular group 
$\Gamma$, which are obtained by imposing that the set 
of all modular transformations be congruent to the identity mod $N$
\begin{equation}
\Gamma(N)=\left \{ 
\left( \begin{matrix}
a & b \\
c & d 
\end{matrix} \right) \subset SL(2, \mathbb{Z}):
\left( \begin{matrix}
a & b \\
c & d 
\end{matrix} \right) = 
\left( \begin{matrix}
1 & 0 \\
0 & 1 
\end{matrix} \right) \, (\mbox{mod} \, N) \right \}.
\end{equation}
In this nomenclature the modular group $\Gamma$ is called the modular group of level 1 and denoted 
as $\Gamma(1)$ \cite{McKean,Diamond}. A relevant mathematical structure is the coset of the modular group with 
the congruence subgroups which are isomorphic to $PSL(2, \mathbb{Z}/N \mathbb{Z})$ \cite{Diamond}
\begin{equation}
\frac{SL(2, \mathbb{Z})}{\Gamma(N)} \rightarrow PSL(2, \mathbb{Z}/N \mathbb{Z}).
\end{equation}

\begin{figure}[htb] 
\center{\includegraphics[height=5cm,width=4.0cm]{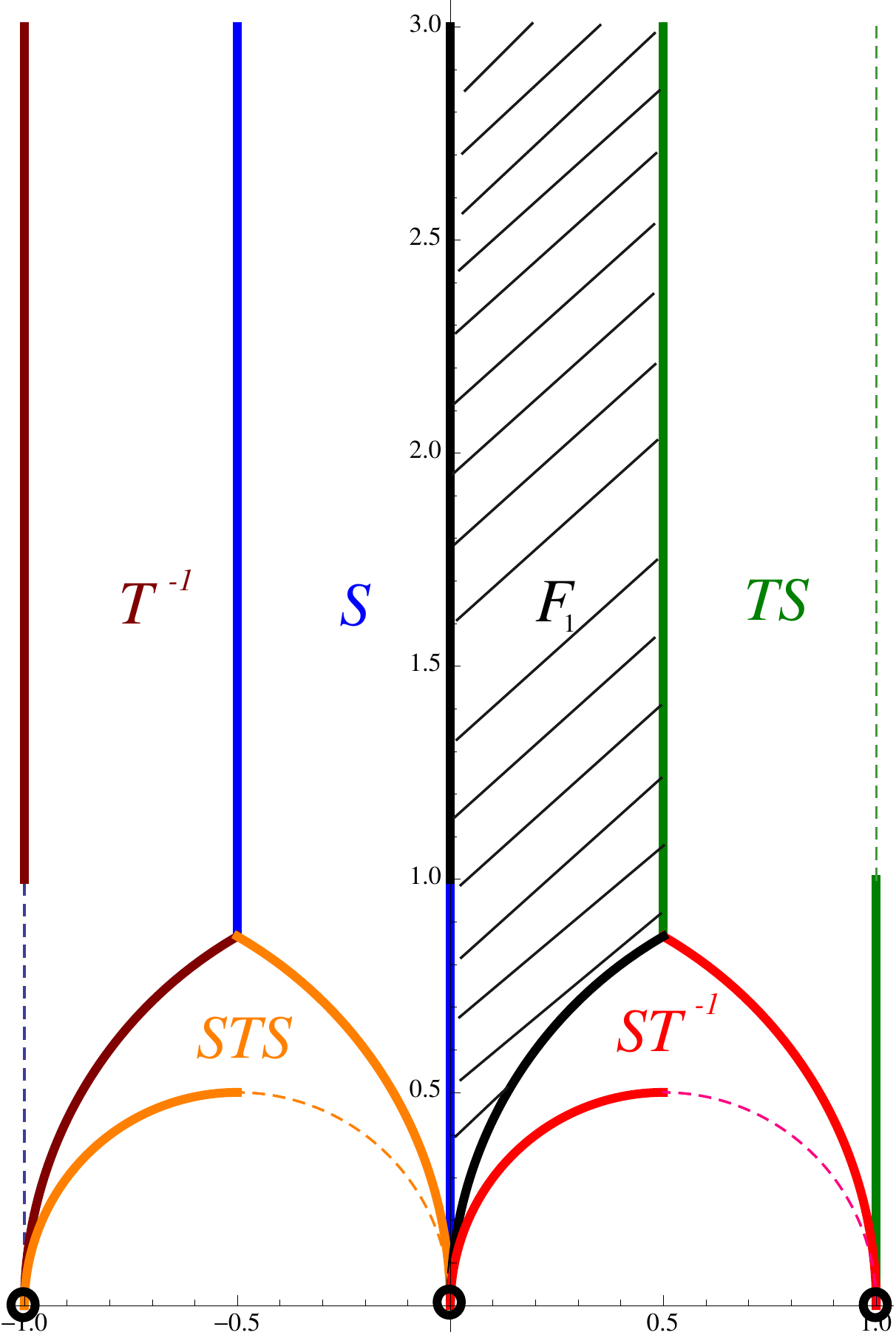}}
\caption{Fundamental cell ${\cal F}_2$ for $\Gamma(2)$. The heavy part of the figure is retained, the rest is not. 
In particular the cusps $-1$, $0$, $1$ and $i \infty$ are excluded.  } \label{RegionG2}
\end{figure}

For the solutions of the  simple pendulum the relevant congruence subgroup is the one of level 2: $\Gamma(2)$. 
It turns out that all the groups $PSL(2, \mathbb{Z}/N \mathbb{Z})$ 
are of finite order and in particular $PSL(2, \mathbb{Z}/2 \mathbb{Z})$ is of order six. In table
\ref{GroupElements} we give explicitly the six elements of the coset and their corresponding form as 
group elements of $PSL(2, \mathbb{Z})$. Analogously to the case of 
the modular group, a fundamental cell for a subgroup $\Gamma(N)$  is a region ${\cal F}_N$ in the 
upper half plane that meets each orbit of $\Gamma(N)$ in a single point. Because $\Gamma(2)$ is of order 
six in $\Gamma$, a fundamental cell for $\Gamma(2)$ can be formed from the six copies of any fundamental cell 
${\cal F}$ of $\Gamma$ produced by the action of the six elements. In figure \ref{RegionG2} 
we show the  fundamental region ${\cal F}_2$ of $\Gamma(2)$. This cell can be obtained 
from the region denoted as ${\cal F}_1$ which is a different fundamental region for $\Gamma$ as compared to 
the usual region ${\cal F}$ of the figure \ref{FundamentalDomain}. ${\cal F}_1$ is obtained if  ${\cal F}$ is replaced 
by its right half, plus inversion of its left half by the $S$ transformation. Thus ${\cal F}_1$ 
consists of the open region
$\{ z: 0 < \mbox{Re}(z) < 1/2 \cap \frac{z \bar{z}}{z+\bar{z}} >1 \}$ and part of its boundary must be included. 
Geometrically $ \frac{z \bar{z}}{z+\bar{z}} =1$ represents a unitary circle with center at $z=1$. 
A possible choice of the boundary includes the set of all points $\{ z: z=iy$ with $y \geq 1\}$ union 
$\{z: \frac{z \bar{z}}{z+\bar{z}} =1$ with 
$0 < $ Re$(z) \leq 1/2\}$. The full fundamental region ${\cal F}_2$  so produced is the part of the half-plane above the 
two circles  of radius $1/2$ centered at $\pm 1/2$. 

As a complementary comment we mention that sometimes in the literature $\Gamma(2)$ appears under the 
name of {\it modular group} $\Lambda$. It turns out that the group is  isomorphic to the symmetric group 
$S_3$, which is  the group of all permutations of a three-element set and also to the dihedral group of 
order six (degree three) $D_6$, which represents, the group of symmetries (rotations and reflections) of 
the equilateral triangle.

\subsection{Lattices}\label{lattices}

A {\it lattice} $L$ is an aggregate of complex numbers 
with two properties  \cite{DuVal}: (i) is a group with respect to addition and (ii) the 
absolute magnitudes of the non-zero elements are bounded below.  
Because the Jacobi elliptic functions are meromorphic functions on  $\mathbb{C}$, that are periodic in two 
directions: $f(z)=f(z+\omega_1)=f(z+\omega_2)$, we are interested in the so-called {\it double lattices}, 
consisting of all linear combinations with integer coefficients of two {\it generating coefficients} 
or {\it primitive periods} $\omega_1, \omega_2 \in \mathbb{C}$, whose ratio is imaginary
\begin{equation}
L(\omega_1,\omega_2) = \{ n\omega_1+m\omega_2 | n,m \in \mathbb{Z}\} \hspace{0.5cm} 
\mbox{such that} \hspace{0.5cm} f(z)=f( z+n\omega_1+m\omega_2 ),\hspace{0.5cm} \forall \, \, z \in \mathbb{C}. 
\end{equation}
The lattice points are the vertices of a pattern of parallelograms filling the whole plane, 
whose sides can be taken to be any pair of generators.
The shapes of the lattices define equivalence classes.  If $L(\omega_1, \omega_2)$ is any lattice, 
and the number $k\neq 0 \in \mathbb{C}$, then $kL(\omega_1, \omega_2)$ denotes the 
aggregate of complex numbers $kz$ for all $z \in L(\omega_1, \omega_2)$ and it is also a lattice, which is said 
to be in the same equivalence class as $L(\omega_1, \omega_2)$. If $\bar{L}$ denotes the aggregate of 
complex numbers $\bar{z}$, $\forall$ $z \in L$; $\bar{L}$ is also a lattice. If $\bar{L}= L$, the lattice is 
called {\it real}. If the primitive periods can be chosen so that $\omega_1$ is real and $\omega_2$ pure 
imaginary, $L$ is called {\it rectangular}. 

Rectangular lattices are real, and they are called horizontal or vertical, according as the longer sides of the 
rectangles are horizontal or vertical. The particular case in which both sides are equal is called the 
{\it square lattice}. Every lattice satisfies $L=-L$, and the only square lattice for which, $L= \alpha L$, with 
$\alpha \neq \pm 1$, is the lattice $iL$. If $L$ is a vertical rectangular lattice, $iL$ is a horizontal rectangular 
lattice and vice versa. 

Associated to the lattice is the concept of {\it residue classes}. If $z$ is any complex variable, $z+L$ denotes 
the aggregate of values $z+\omega$ for all $\omega$ in the lattice $L$. This aggregate is called a residue class 
(mod $L$).  The residue classes (mod $L$) form a continuous group under addition, defined in the way
$(z+L) +(w +L)= (z+w) +L$. $L$ itself is a residue class (mod $L$), the zero element of the group.
These residue classes allow to introduce the concept of {\it fundamental region} of $L$, consisting in a 
simply connected region of the complex plane which contains exactly one member  of each residue 
class (mod $L$) \footnote{Be aware that we are following the mathematics literature in which often it is 
referred with the same name {\it fundamental region},  to two different kind of 
regions, the one that we have denoted as ${\cal F}_N$ and the one just described. We expect do not 
generate confusion.}.
A fundamental region can be chosen in many ways, the simplest and usually the most 
convenient, is what is called either a  {\it unit cell}, a {\it fundamental cell} or a {\it fundamental parallelogram}  
which is defined by all the points of the sides $\vec{\omega}_1$ and $\vec{\omega_2}$, including the 
vertex $\vec{0}$, but excluding the rest of the boundary and of course the whole interior points of the 
parallelogram. Mathematically the cell is given by the coset space $\mathbb{C}/L(\omega_1,\omega_2)$, 
where abusing of the notation, in this expression $L$ is considered as a residue class.
Since the opposite sides of the fundamental cell must be identified, the coset space 
$\mathbb{C}/L(\omega_1,\omega_2)$ is homeomorphic to the torus $\mathbb{T}^2$. In other words, the pair 
$(\omega_1, \omega_2)$ defines a complex structure of $\mathbb{T}^2$ \cite{Nakahara:2003nw}. 

The shape of the lattice is determined by the {\it modular parameter} $\tau \equiv \omega_2/\omega_1$. 
It is important to note that, while a pair of primitive periods $\omega_1$, $\omega_2$, 
generates a lattice, a lattice does not have any unique pair of primitive periods, that is, many 
fundamental pairs (in fact, an infinite number)  correspond to the same lattice. Specifically a change of 
generators $\omega_1$, $\omega_2$ to $\omega'_1$ and $\omega'_2$ of the form 
\begin{equation}\label{ChangePeriods}
\left( \begin{matrix}
\omega'_2 \\ \omega'_1
\end{matrix} \right) =
\left( \begin{matrix}
a & b \\
c & d 
\end{matrix} \right)
\left( \begin{matrix}
\omega_2 \\ \omega_1
\end{matrix} \right),
\end{equation}
induces a mapping on the modular parameter $\tau$, belonging to the modular group. 
These maps are the link between the concepts of lattices, torus and modular group. 
As an example we discuss 
the mapping on $\tau$ induced by the generators (\ref{Generators}). The generator $S$ interchanges the roles 
of the generators of the lattice $\omega_1 \leftrightarrow \omega_2$ or equivalently it changes the longitude 
$l$ for the meridian $m$ of the torus and vice versa. The transformation $T$ generates a {\it Dehn twist} along the 
meridian which can be understood as follows \cite{Nakahara:2003nw}.  As a first step cut the torus along the 
meridian $m$, then take one of the lips of the cut and rotate it by $2 \pi$ with the other lip kept fix and finally 
glue the lips together again.

\begin{figure}[htb] 
\center{\includegraphics[height=3cm,width=9.0cm]{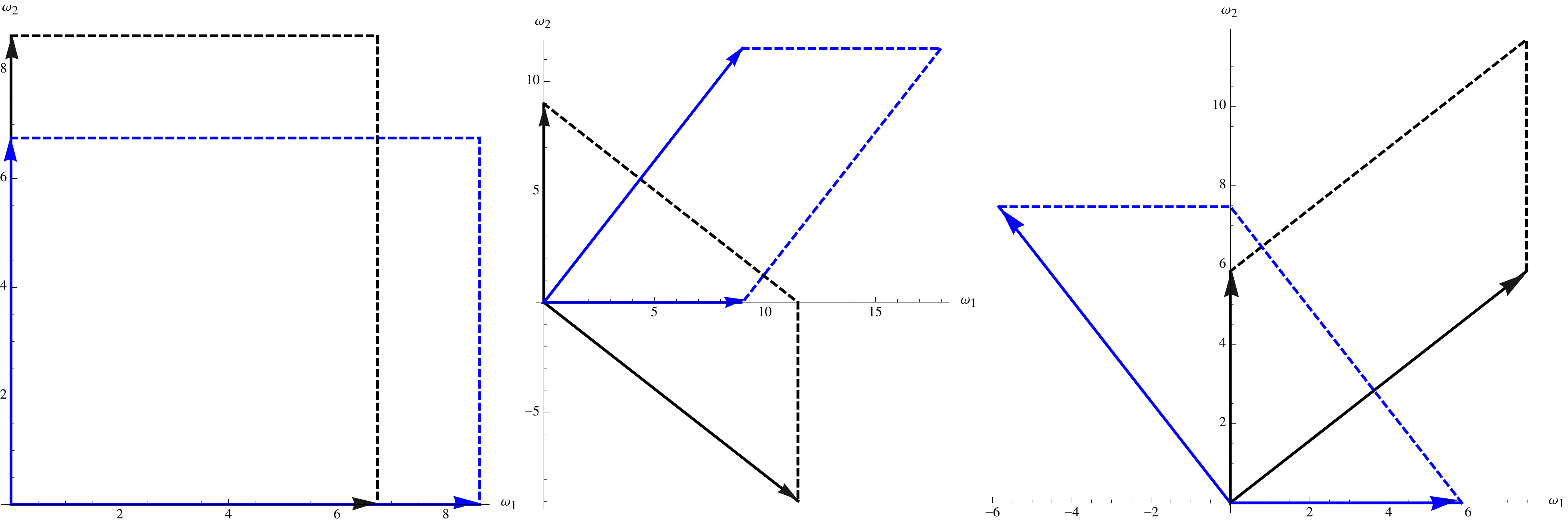}}
\caption{Figure shows the fundamental cell for six different normal lattices. In the first plot the vertical
cell corresponds to a value of the square modulus $k^2=1/4$ and belongs to a normal lattice of the 
type $L^*$. Under an 
$S$ transformation, the cell transforms to the horizontal one whose value of the square modulus is $k^2=3/4$ and 
belongs to a normal lattice $iL^*$. Analogously in the second and third plots, the fundamental cells in black 
belong to the lattices $kL^*$ and $ik_cL^*$ respectively, with 
values of the square modulus $k^2=4$ and $k^2=4/3$. The blue cells are 
obtained as their $S$-dual fundamental cells and have the values $k^2=-3$ and $k^2=-1/3$
and belong to normal lattices of the kind $ikL^*$ and $k_cL^*$ respectively. In every case, the continuous 
lines are included in the fundamental cell, whereas the dashed lines are not. The numerical values of the 
two generators $\omega_1$ and $\omega_2$ are given in table \ref{Numerica}.} \label{CellS}
\end{figure}


If the stationary values $e_1$, $e_2$ and $e_3$ are the roots of the cubic equation $4x^3-g_2x-g_3=0$,
for any lattice $L$, with assigned generators $\omega_1$, $\omega_2$, we can define the scale constant $h$ 
by means of the relation: $h^2=e_1-e_2$, and the moduli as
\begin{equation}
k^2 = \frac{e_3-e_2}{h^2}, \hspace{1cm}  k_c^2 = \frac{e_1-e_3}{h^2}.
\end{equation}
A lattice for which $h^2=1$ is called {\it normal}, and using the notation  of \cite{DuVal}, we write it with 
a star $L^*$. Every lattice $L$ with assigned generators is similar to a unique normal lattice $L^* = hL$ with 
corresponding generators, since $e_i(hL)=h^{-2}e_i(L)$. For a given lattice shape with no assignments of 
generators, there are {\it six} normal lattices, as any of the six differences $e_i-e_j$ can be 
taken as $h^2$. If one of these is $L^*$, with modulus $k$, the others are 
$iL^*$, $kL^*$, $ikL^*$, $k_c L^*$ and $ik_c L^*$, with moduli $k_c$, 
$1/k$, $ik_c/k$, $ik/k_c$ and $1/k_c$  respectively, where $k^2+k_c^2=1$. These fall into three pairs which are 
of the same size, interchanged by a rotation of a right angle.

For the rectangular lattice shape, the six normal lattices are all real. Ordinarily $\omega_1$ is taken real and 
$\omega_2$ pure imaginary, so that $e_1>e_3>e_2$, and $0<k^2<1$, $0<k_c^2<1$, with $k^2<1/2<k_c^2$ 
if $L$ is vertical. We summarize the properties of the normal lattices in table \ref{GroupElements}

\begin{table}[htb]  
\begin{tabular}{|c | c | c | c | c | c |c|} 
\hline
$\Gamma$ & $PSL(2, \mathbb{Z})$ & $PSL(2, \mathbb{Z}/2 \mathbb{Z})$ & Modulus &
Quarter periods & Action on $k^2$ &Normal lattice \\
\hline 
\hline
 $\tau$ & $\pm \mathbb{I}$ & $ \pm \left( \begin{array}{cc}
1 & 0 \\ 0 & 1 
\end{array} \right) $ & $k$& $K$,  $iK_c$ & $k^2 \in(0,1/2] $ & $ L^*$\\
$- \frac{1}{\tau}$ & $\pm S$ & $ \pm \left( \begin{array}{cc} 
0 & -1 \\ 1 & \, \, \, \, 0 
\end{array} \right) $ & $k_c$ &  $K_c$,  $iK$ & $1-k^2 \in [1/2,1)$ & $i L^*$ \\
$\frac{\tau}{1-\tau}$ & $\pm STS$ & $ \mp \left( \begin{array}{cc}
\, \, \, \, 1 & 0 \\ -1 & 1 
\end{array} \right) $ &  $\frac{1}{k}$ & $k(K-iK_c)$, $ikK_c$ & 
$\frac{1}{k^2} \in [2,\infty)$ &$k L^*$\\
$\frac{\tau - 1}{\tau}$ & $\pm TS$ &  $\pm  \left( \begin{array}{cc}
1 & -1 \\ 1 & \, \, \, \, 0 
\end{array} \right) $ & $i\frac{k_c}{k}$ & $kK_c$, $ik(K-iK_c)$  &  
$1-\frac{1}{k^2} \in (-\infty, -1]$ &$i k L^*$ \\ 
$\frac{1}{1-\tau}$ & $\pm ST^{-1}$ & $ \mp  \left( \begin{array}{cc}
 \, \, \, \, 0 & 1 \\ -1 & 1 
\end{array} \right) $ & $\frac{1}{k_c}$ &  $k_c(K_c+iK)$, $ik_cK$ & 
$\frac{1}{1-k^2} \in (1,2] $& $ i  k_c L^*$ \\
$\tau-1$ & $\pm T^{-1}$ & $ \pm \left( \begin{array}{cc}
1 & -1 \\ 0 & \, \, \, \, 1 
\end{array} \right) $ & $i \frac{k}{k_c}$ & $k_cK$, $ik_c(K_c+iK)$& 
$\frac{k^2}{k^2-1} \in [-1,0)$ & $  k_c L^*$ \\
\hline
\end{tabular}
\caption{Main characteristics of the six order $PSL(2, \mathbb{Z}/2 \mathbb{Z})$ group and its relation 
to the six normal lattices.}\label{GroupElements}
\end{table}


\section{Jacobi elliptic functions}\label{Jacobi}

In the previous appendix we reviewed the action of the modular group on the modular parameter. In this 
appendix we want to specialize that discussion to the case of the elliptic Jacobi functions. In particular we 
are interested in the relation between  the  six dimensional group $PSL(2, \mathbb{Z}/2 \mathbb{Z})$
and what is called transformations of the elliptic Jacobi functions. There are three transformations that are 
exposed often in the literature, the {\it  Jacobi's imaginary transformation}, the {\it Jacobi's imaginary modulus 
transformation} and the {\it Jacobi's real transformation}. These are transformations that relate the Jacobi elliptic 
functions with different value of the square modulus $k^2$.  Behind these transformations is the property
that the modulus of the Jacobi functions can be defined in the real line $k^2 \in \mathbb{R}$ with 
exception of the points $z=-1, 0, 1$, and it can be divided in six intervals
\[
k^2 \in (-\infty, -1] \, \cup [-1, 0) \cup (0,1/2] \cup [1/2,1) \cup (1,2] \cup [2,\infty) .
\]
These six intervals are in one to one relation to the column {\it Action on} $k^2$ in table (\ref{GroupElements}), if 
we consider that the modulus in the fundamental region ${\cal F}_1$ of $PSL(2,\mathbb{Z})$ takes values in the 
interval $0 \leq k^2 \leq 1/2$. In the following we summarize some of the properties of the Jacobi elliptic
functions that are useful throughout the paper.

\subsection{Jacobi elliptic functions with modulus $0<k^2<1$}\label{BasicJacobi}

The Jacobi elliptic functions are meromorphic functions in ${\mathbb C}$, that have a fundamental real period and a 
fundamental complex period, i.e., they are doubly periodic. The periods are determined by the value of the square 
modulus and in the following we assume that $0<k^2<1$.

The primitive real period of the three basic functions can be inferred from the following relations which 
are dictated by the addition formulas for the Jacobi functions 
\cite{Whittaker1927,DuVal,Lang,Lawden,McKean,Armitage}
\begin{equation}\label{BasicRelEK}
\mbox{sn}(z+K,k) = \frac{\mbox{cn}(z,k)}{\mbox{dn}(z,k)}, \hspace{0.5cm} 
\mbox{cn}(z+K,k) = -k_c \frac{\mbox{sn}(z,k)}{\mbox{dn}(z,k)}, \hspace{0.5cm} 
\mbox{dn}(z+K,k) =k_c  \frac{1}{\mbox{dn}(z,k)},
\end{equation}
where the quarter-period $K$ is defined as function of the square modulus $k^2$ as
\begin{equation}\label{DefK}
K \equiv \int_0^1 \frac{du}{\sqrt{(1-u^2)(1-k^2 u^2)}}.
\end{equation}
In particular we obtain the values \mbox{sn}$(K,k) =1$, \mbox{cn}$(K,k)=0$ and \mbox{dn}$(K,k) = k_c$, from 
the ones \mbox{sn}$(0,k)=0$, \mbox{cn}$(0,k)=1$ and \mbox{dn}$(0,k) = 1$. 
Iteration of relations (\ref{BasicRelEK}) leads to
\begin{equation}\label{BasicRelE2K}
\mbox{sn}(z+2K,k) = - \mbox{sn}(z,k), \hspace{0.5cm} 
\mbox{cn}(z+2K,k) = - \mbox{cn}(z,k), \hspace{0.5cm} 
\mbox{dn}(z+2K,k) = \mbox{dn}(z,k).
\end{equation}
The last relation is telling that the function dn($z,k$) has real period $2K$. A further $2K$ iteration will tell us that 
the other two Jacobi elliptic functions (sn($z,k$) and cn($z,k$)) have primitive real period $4K$.
Regarding the complex period, we have the relations
\begin{equation}\label{RelJacobiKc}
\mbox{sn}(z+iK_c,k) = \frac{1}{k}\frac{1}{\mbox{sn}(z,k)}, \hspace{0.5cm} 
\mbox{cn}(z+iK_c,k) = -i\frac{1}{k} \frac{\mbox{dn}(z,k)}{\mbox{sn}(z,k)}, \hspace{0.5cm} 
\mbox{dn}(z+iK_c,k) = -i \frac{\mbox{cn}(z,k)}{\mbox{sn}(z,k)},
\end{equation}
where $K_c$ is defined as function of  the so-called complementary modulus $k_c^2\equiv 1 - k^2$ in 
the form
\begin{equation}\label{DefKc}
K_c \equiv \int_0^1 \frac{du}{\sqrt{(1-u^2)(1-k_c^2 u^2)}}.
\end{equation}
Iterating these relations once leads to
\begin{equation}\label{BasicRelE2iK}
\mbox{sn}(z+2iK_c,k) = \mbox{sn}(z,k), \hspace{0.5cm} 
\mbox{cn}(z+2iK_c,k) = - \mbox{cn}(z,k), \hspace{0.5cm} 
\mbox{dn}(z+2iK_c,k) = - \mbox{dn}(z,k).
\end{equation}
The first relation is telling us that the elliptic function sn($z,k$) has a pure imaginary primitive period $2iK_c$. A 
further $2iK_c$ iteration leads to the conclusion that the elliptic function dn($z,k$) has a pure imaginary 
primitive period $4iK_c$ whereas the elliptic function cn($z,k$) has a fundamental period $4iK_c$. In the latter case 
notice that combining the second relation of (\ref{BasicRelE2K}) and the second relation of (\ref{BasicRelE2iK}) 
leads to the result cn$(z+2K+2iK_c,k)=\mbox{cn}(z,k)$ concluding that this elliptic function has a primitive complex 
period $2K+2iK_c$. In summary, the primitive periods of the three basic Jacobi functions are 
\begin{eqnarray}
\mbox{sn}(z,k) &=& \mbox{sn}(z+4K,k)= \mbox{sn}(z+2iK_c,k),\\
\mbox{cn}(z,k) &=&\mbox{cn}(z+4K,k)= \mbox{cn}(z+2K+2iK_c,k),\\
\mbox{dn}(z,k) &=& \mbox{dn}(z+2K,k)=\mbox{dn}(z+4iK_c,k).
\end{eqnarray} 
Because these periods do not coincide one looks for two common periods  in order to define a common 
fundamental cell for the three functions. 
These {\it fundamental periods} are $4K$ and  $4iK_c$, they are not primitive because linear 
combinations of them does not give origin for instance to the primitive period $2K+2iK_c$ of cn$(z,k)$. 
The fundamental cell for the Jacobi elliptic functions is, therefore, the parallelogram with vertices 
$(0, 4K, 4iK_c, 4K+4iK_c)$, and the modular parameter $\tau$ turns out  to be
\begin{equation}\label{DefTEllip}
\tau \equiv  \frac{iK_c}{K}.
\end{equation}
Given this definition of the modular parameter we see that not every point of  ${\cal F}_1$ 
corresponds to a modulus 
$k^2$ of the Jacobi functions but only the values on the vertical boundary $\tau \in [i,i \infty)$, 
being the point $\tau=i$ the one that corresponds to $k^2=1/2$, since in this case $K=K_c$ and therefore the 
corresponding normal lattice is squared. The rest of points on the vertical boundary corresponds to vertical 
normal lattices because $K<K_c$ and all of them have a value of the square modulus $0<k^2<1/2$. By acting 
the five group elements of $PSL(2, \mathbb{Z}/2 \mathbb{Z})$ different from the identity to the 
modular parameter values on the vertical boundary of ${\cal F}_1$, we can generate the whole set of possible 
values of $\tau$ and therefore the whole set of possible values of the square modulus $k^2$ of the Jacobi 
functions (see figure \ref{LatCel}).  

Derivatives of the basic functions, which are necessary to obtain the angular velocities are
\begin{equation}\label{BasDer}
\frac{d}{dz} \mbox{sn}(z,k) = \mbox{cn} (z,k) \, \mbox{dn}(z,k), \hspace{0.3cm} 
\frac{d}{dz} \mbox{cn}(z,k) = - \mbox{sn} (z,k) \, \mbox{dn}(z,k), \hspace{0.3cm} 
\frac{d}{dz} \mbox{dn}(z,k) = - k^2 \mbox{sn} (z,k) \, \mbox{cn}(z,k).
\end{equation}

\subsection{Jacobi's imaginary transformation}

The transformation induced by the generator $S(\tau)$  of the modular group on the Jacobi functions with modulus 
$k$, is known as the Jacobi's imaginary transformation.  In this case 
the modulus and the complementary modulus exchange with each other
\begin{equation}
k \mapsto k_c \hspace{0.5cm}  \mbox{and}  \hspace{0.5cm}  k_c \mapsto k \hspace{0.5cm} 
\Rightarrow \hspace{0.5cm} K \mapsto K_c  \hspace{0.5cm}  \mbox{and}  \hspace{0.5cm} K_c \mapsto K.  
\end{equation}
Applying this transformation on the vertical boundary of ${\cal F}_1$, generates both the transformed pure 
imaginary modular parameter and the transformed modulus which belong to the intervals $\tau \in (0,i]$ and 
$1/2 \leq k_c^2=1- k^2 < 1$, respectively. The Jacobi functions itself transform as 
\begin{equation}\label{TransJacS}
\mbox{sn}(z, k_c) = -i \, \mbox{sc} (iz,k), \hspace{0.5cm}
\mbox{cn}(z, k_c) = \mbox{nc} (iz,k), \hspace{0.5cm}
\mbox{dn}(z, k_c) = \mbox{dc} (iz,k).
\end{equation}
This is the mathematical property behind the analysis made by Appell to deal with solutions of imaginary time. 
These transformations are used very often to change a pure imaginary argument $ix$ to one real $x$, obtaining 
\begin{equation}
\mbox{sn}(ix, k) = i \, \mbox{sc} (x,k_c),  \hspace{0.5cm}
\mbox{cn}(ix, k) = \mbox{nc} (x,k_c),  \hspace{0.5cm}
\mbox{dn}(ix, k) = \mbox{dc} (x,k_c). 
\end{equation}
From a geometrical point of view the normal vertical cell $L^* $ with vertices 
$(0, 4K, 4iK_c,4K+4iK_c)$ changes to the normal horizontal cell  $iL^*$ with vertices 
$(0,4K_c, 4iK, 4K_c+4iK)$ and the corresponding torus 
is obtained from the original one by an interchange of their respective meridians and longitudes. The rest of 
properties of the functions are obtained from the ones in (section \ref{BasicJacobi}) by setting $z=ix$ and 
implementing in the expressions the interchanges $k \leftrightarrow k_c$ and $K \leftrightarrow K_c$.

\subsection{Jacobi's imaginary modulus transformation}

The transformation  induced by the generator $T(\tau)$ of the modular group on the Jacobi functions, is 
known as the imaginary modulus transformation, because under this transformation the modulus change as
\begin{equation}\label{ImagModulus}
k \mapsto i \frac{k}{k_c},  \hspace{0.5cm}  \mbox{and}  \hspace{0.5cm}  k_c \mapsto \frac{1}{k_c},
\end{equation}
which induces a change in the quarter periods of the form
\begin{equation}\label{ImagQuarter}
K \mapsto k_c K,  \hspace{0.5cm}  \mbox{and}  \hspace{0.5cm} K_c \mapsto   k_c(K_c-iK) .  
\end{equation}
Applying this transformation to the vertical boundary of ${\cal F}_1$, generates the transformed modular parameter 
which lies on the vertical line $\tau \in [1+i, 1+i \infty)$ 
and the transformed square modulus which takes values in the interval $-1 \leq  \frac{k^2}{k^2-1} < 0$.
The transformation rule for the Jacobi functions itself are
\begin{equation}\label{CambioT}
\mbox{sn}(z, ik/k_c) = k_c  \, \mbox{sd} (z/k_c,k),  \hspace{0.5cm}
\mbox{cn}(z, ik/k_c) = \mbox{cd} (z/k_c,k), \hspace{0.5cm}
\mbox{dn}(z, ik/k_c) = \mbox{nd} (z/k_c,k).
\end{equation}
It is clear that this transformation allows us to define the Jacobi functions with an imaginary modulus in terms 
of the Jacobi functions with real modulus. Replacing $z \mapsto k_c z$, we can express these transformations 
in its more usual form
\begin{equation}\label{CambioTUsual}
\mbox{sn}(k_cz, ik/k_c) = k_c  \, \mbox{sd} (z,k), \hspace{0.5cm}
\mbox{cn}(k_cz, ik/k_c) = \mbox{cd} (z,k),  \hspace{0.5cm}
\mbox{dn}(k_cz, ik/k_c) = \mbox{nd} (z,k).
\end{equation}
From a geometrical point of view the fundamental vertical cell with vertices $(0, 4K, 4iK_c,4K+4iK_c)$ changes to 
the fundamental cell  with vertices $(0,4k_cK, 4k_cK+4ik_cK, 8k_cK+4ik_cK_c)$ and the corresponding torus 
is changed by a Dehn twist.
Notice that by applying further the transformation $S$ to these expressions, we obtain a fundamental cell  
where the quarter periods (\ref{ImagQuarter}) are interchange among them an the value 
of the square modulus is defined in the interval,  $1 < \frac{1}{1-k^2} \leq 2$, since the modulus 
(\ref{ImagModulus}) also interchanges one to the another.

The elliptic Jacobi functions with negative square modulus satisfy analogous relations to the Jacobi functions
with modulus $0<k^2<1$, these are obtained from equations (\ref{CambioT}) and the corresponding 
relation of the Jacobi functions with  $0<k^2<1$. For instance, the equations analogous to (\ref{relJac1}) and 
(\ref{relJac2}) are
\begin{equation}\label{ksdmodim}
\mbox{sn}^2(z,ik/k_c) + \mbox{cn}^2(z,ik/k_c)=1, 
\hspace{0.3cm} \mbox{and} \hspace{0.3cm}
-\frac{k^2}{k_c^2}\mbox{sn}^2(z,ik/k_c) + \mbox{dn}^2(z,ik/k_c)=1.
\end{equation}
Proceeding in a similar way it is possible  to obtain the equations analogous to (\ref{BasicRelEK}), these are
\begin{equation}
\mbox{sn}(z+K,ik/k_c) = \frac{\mbox{cn}(z,ik/k_c)}{\mbox{dn}(k_cz,ik/k_c)}, \hspace{0.1cm}
\mbox{cn}(z+K,ik/k_c) = -\frac{1}{k_c} \frac{\mbox{sn}(z,ik/k_c)}{\mbox{dn}(k_cz,ik/k_c)}, \hspace{0.1cm}
\mbox{dn}(z+K,ik/k_c) = \frac{1}{ k_c}  \frac{1}{\mbox{dn}(z,ik/k_c)}, 
\end{equation}
which iterating once lead to the relations
\begin{equation}
\mbox{sn}(z+2K,ik/k_c) =-\mbox{sn}(z,ik/k_c), \hspace{0.2cm}
\mbox{cn}(z+2K,ik/k_c) =-\mbox{cn}(z,ik/k_c), \hspace{0.2cm}
\mbox{dn}(z+2K,ik/k_c) =\mbox{dn}(z,ik/k_c).
\end{equation}
The third relation is telling us that the function dn($z,ik/k_c$) has a fundamental period $2K$. A further $2K$ 
iteration leads to the conclusion that the other two Jacobi functions have a fundamental period of $4K$.
Regarding the imaginary period, the equations analogous to (\ref{RelJacobiKc}) are
\begin{eqnarray}
\mbox{sn}(z+iK_c,ik/k_c) &=& i\frac{k_c}{k} \frac{\mbox{dn}(z,ik/k_c)}{\mbox{cn}(k_cz,ik/k_c)}, \\
\mbox{cn}(z+iK_c,ik/k_c) &=& \frac{1}{k} \frac{1}{\mbox{cn}(z,ik/k_c)}, \\
\mbox{dn}(z+iK_c,ik/k_c) &=& i \frac{1}{ k_c}  \frac{\mbox{sn}(z,ik/k_c)}{\mbox{cn}(k_cz,ik/k_c)},
\end{eqnarray}
which after an iteration lead to
\begin{equation}
\mbox{sn}(z+2iK_c,ik/k_c) =-\mbox{sn}(z,ik/k_c), \hspace{0.1cm}
\mbox{cn}(z+2iK_c,ik/k_c) =\mbox{cn}(z,ik/k_c), \hspace{0.1cm}
\mbox{dn}(z+2iK_c,ik/k_c) =-\mbox{dn}(z,ik/k_c).
\end{equation}
These relations indicate that the function cn$(z, ik/k_c)$ has fundamental imaginary period $2iK_c$, whereas the 
other two Jacobi functions have $4iK_c$. In summary, the primitive periods of the three basic Jacobi functions are 
\begin{eqnarray}
\mbox{sn}(z,ik/k_c) &=& \mbox{sn}(z+4K,ik/k_c)= \mbox{sn}(z+2K+2iK_c,ik/k_c),\\
\mbox{cn}(z,ik/k_c) &=&\mbox{cn}(z+4K,ik/k_c)= \mbox{cn}(z+2iK_c,ik/k_c),\\
\mbox{dn}(z,ik/k_c) &=& \mbox{dn}(z+2K,ik/k_c)=\mbox{dn}(z+4iK_c,ik/k_c).
\end{eqnarray} 
 It is straightforward to verify that in this case the derivatives of the fundamental relations that follows from 
(\ref{BasDer}) are
\begin{equation}
\frac{d}{dz} \mbox{sn}(z,ik/k_c) = \mbox{cn} (z,ik/k_c) \, \mbox{dn}(z,ik/k_c), \hspace{0.3cm} 
\frac{d}{dz} \mbox{cn}(z,ik/k_c) = - \mbox{sn} (z,ik/k_c) \, \mbox{dn}(z,ik/k_c), 
\end{equation}
and
\begin{equation}
\frac{d}{dz} \mbox{dn}(z,ik/k_c) = \frac{ k^2}{k_c^2} \mbox{sn} (z,ik/k_c) \, \mbox{cn}(z,ik/k_c).
\end{equation}

\subsection{Jacobi's real transformation}\label{JacobiSTS}

In the literature of the elliptic functions, the transformation generated by the element $STS$ of the modular group 
\begin{equation}
\tau_I=\frac{\tau}{1-\tau},
\end{equation}
which can be obtained as a composition of the following three transformations 
\begin{equation}
\tau_I=-\frac{1}{\tau_2}, \hspace{1cm}  \tau_2=1+\tau_1, \hspace{1cm} \mbox{and}  \hspace{1cm} 
\tau_1=-\frac{1}{\tau},
\end{equation}
generates the so-called Jacobi's real transformation. Under it, the modulus of the elliptic functions change as
\begin{equation}\label{ModulusReal}
k \mapsto \frac{1}{k},  \hspace{0.5cm}  \mbox{and}  \hspace{0.5cm}  k_c \mapsto i \frac{k_c}{k},
\end{equation}
whereas the quarter periods transform as
\begin{equation}\label{QuarterReal}
K \mapsto k(K-iK_c),  \hspace{0.5cm}  \mbox{and}  \hspace{0.5cm} K_c \mapsto  k K_c .  
\end{equation}
Applying this transformation to the vertical boundary of ${\cal F}_1$, generates the transformed modular parameter 
which lies on the line $\tau  = -\frac{y^2}{1+y^2} + i \frac{y}{1+y^2}$, with $y$ in the interval $y \in [1,\infty)$  and the 
transformed square modulus which takes values in the interval $ 2 \leq 1/k^2 < \infty$. The transformation rules 
for the Jacobi functions itself are
\begin{equation}\label{UsefulSTS}
\mbox{sn}(z, 1/k) = k\,   \mbox{sn} (z/k,k), \hspace{0.5cm}
\mbox{cn}(z, 1/k) = \mbox{dn}  (z/k,k), \hspace{0.5cm}
\mbox{dn}(z, 1/k) = \mbox{cn}  (z/k,k). 
\end{equation}
This transformations allows us to define the Jacobi elliptic functions with square modulus greater than two in terms 
of Jacobi functions with modulus $0<k^2 \leq1/2$. Replacing $z \mapsto k z$, allows to express these 
transformations in its more usual form
\begin{equation}
\mbox{sn}(k z, 1/k) = k\,   \mbox{sn} (z,k), \hspace{0.5cm}
\mbox{cn}(k z, 1/k) = \mbox{dn}  (z,k),\hspace{0.5cm}
\mbox{dn}(k z, 1/k) = \mbox{cn}  (z,k).
\end{equation}
A further application of the group transformation $S$ to these expressions leads to the interchange of the modulus 
(\ref{ModulusReal}) and to the interchange of the quarter periods (\ref{QuarterReal}). In this case the square 
modulus of the Jacobi functions is defined in the interval $-\infty < 1-1/k^2 \leq -1$.

As in the previous cases it is possible to obtain the fundamental periods of the three different basic Jacobi elliptic 
functions, since the arguments as before, we just list the equations. For the real period we have
\begin{equation}
\mbox{sn}(z+K, 1/k) = k\,  \frac{\mbox{dn}(z,1/k)}{\mbox{cn}(z,1/k)}, \hspace{0.3cm}
\mbox{cn}(z+K, 1/k) = k_c \,  \frac{1}{\mbox{cn}(z,1/k)}, \hspace{0.3cm}
\mbox{dn}(z+K, 1/k) = -\frac{k_c}{k} \,  \frac{\mbox{sn}(z,1/k)}{\mbox{cn}(z,1/k)}, 
\end{equation}
and iterating we get
\begin{equation}
\mbox{sn}(z+2K, 1/k) = - \mbox{sn}(z,1/k), \hspace{0.3cm}
\mbox{cn}(z+2K, 1/k) = \mbox{cn}(z,1/k), \hspace{0.3cm}
\mbox{dn}(z+2K, 1/k) = - \mbox{dn}(z,1/k).
\end{equation}
As for the imaginary period
\begin{equation}
\mbox{sn}( z+iK_c, 1/k) = \frac{k}{\mbox{sn}(z,1/k)}, \hspace{0.1cm}
\mbox{cn}(z+iK_c, 1/k) = -ik \frac{{\mbox{dn}(z,1/k)}}{\mbox{sn}(z,1/k)}, \hspace{0.1cm}
\mbox{dn}(z+iK_c, 1/k) = -i \frac{\mbox{cn}(z,1/k)}{\mbox{sn}(z,1/k)},
\end{equation}
and after an iteration
\begin{equation}
\mbox{sn}(z+2iK_c, 1/k) = \mbox{sn}(z,1/k), \hspace{0.1cm}
\mbox{cn}(z+2iK_c, 1/k) = - \mbox{cn}(z,1/k) , \hspace{0.1cm}
\mbox{dn}(z+2iK_c, 1/k) = - \mbox{dn}(z,1/k).
\end{equation}
 In summary, the primitive periods of the three basic Jacobi functions are 
\begin{eqnarray}
\mbox{sn}(z,1/k) &=& \mbox{sn}(z+4K,1/k)= \mbox{sn}(z+2iK_c,1/k),\\
\mbox{cn}(z,1/k) &=&\mbox{cn}(z+2K,1/k)= \mbox{cn}(z+4iK_c,1/k),\\
\mbox{dn}(z,1/k) &=& \mbox{dn}(z+4K,1/k)=\mbox{dn}(z+2K+2iK_c,1/k).
\end{eqnarray} 
Finally, the equations analogous to (\ref{relJac1}) and (\ref{relJac2}) are
\begin{equation}\label{ksdmodinv}
\mbox{sn}^2(z,1/k) + \mbox{cn}^2(z,1/k)=1, 
\hspace{0.3cm} \mbox{and} \hspace{0.3cm}
\frac{1}{k^2}\mbox{sn}^2(z,1/k) + \mbox{dn}^2(z,1/k)=1.
\end{equation}
whereas the derivatives of the basic functions are
\begin{equation}
\frac{d}{dz} \mbox{sn}(z,1/k) = \mbox{cn} (z,1/k) \, \mbox{dn}(z,1/k), \hspace{0.3cm} 
\frac{d}{dz} \mbox{cn}(z,1/k) = - \mbox{sn} (z,1/k) \, \mbox{dn}(z,1/k), 
\end{equation}
and 
\begin{equation}
\frac{d}{dz} \mbox{dn}(z,1/k) = - \frac{1}{k^2} \mbox{sn} (z,1/k) \, \mbox{cn}(z,1/k).
\end{equation}

\acknowledgments 
The author would like to thank Manuel de la Cruz, N\'{e}stor Gaspar and Lidia Jim\'{e}nez for 
their valuable comments. This work is  partially supported from 
CONACyT Grant No. 237351 ``Implicaciones f\'{i}sicas de la estructura del espacio-tiempo".

\bibliography{notes}
\end{document}